\documentclass[aps,pra,twocolumn,groupedaddress,showpacs]{revtex4-1}
\usepackage{amsmath,amssymb}
\usepackage[dvipdfm]{graphicx}
\usepackage{natbib}
\usepackage{bm}
\usepackage{braket}

\begin{document}
\title{Finite temperature dynamical properties of SU($N$) fermionic Hubbard
models in the spin-incoherent regime}

\author{Akiyuki Tokuno}
\affiliation{DPMC-MaNEP, University of Geneva, 24 Quai Ernest-Ansermet
CH-1211 Geneva, Switzerland.}

\author{Thierry Giamarchi}
\affiliation{DPMC-MaNEP, University of Geneva, 24 Quai Ernest-Ansermet
CH-1211 Geneva, Switzerland.}

\date{\today}

\begin{abstract}
 We study strongly correlated Hubbard systems extended to symmetric $N$-component
 fermions.
 We focus on the intermediate-temperature regime between magnetic
 superexchange and interaction energy, which is relevant to current
 ultracold fermionic atom experiments.
 The $N$-component fermions are represented by slave particles, and, by
 using a diagrammatic technique based on the atomic limit, spectral functions
 are analytically obtained as a function of temperature, filling factor
 and the component number $N$.
 We also apply this analytical technique to the
 calculation of lattice modulation experiments.
 We compute the production rate of double occupancy induced by modulation of an
 optical lattice potential.
 Furthermore, we extend the analysis to take into account the trapping
 potential by use of the local density approximation. We find an
 excellent agreement with recent experiments on $^{173}$Yb atoms.
\end{abstract}


\maketitle

\section{Introduction}
The realization of ultracold atoms in an optical lattice
opens up the possibility to study in a controlled way strongly
correlated quantum
systems~\cite{Esslinger/AnuuRevCondMattPhys1.2010:review,Bloch.Dalibard.Zwerger/RevModPhys80.2008}.
Such strongly correlated atoms are well described by the Hubbard model.
This model plays a central role for the study of the Mott insulator (MI)
transition~\cite{Imada.Fujimori.Tokura/RevModPhys70.1998:review},
high-$T_c$
superconductivity~\cite{Lee.Nagaosa.Wen/RevModPhys78.2006:review}, and
quantum magnetism.
In addition, the high controllability of model parameters such as
the interaction by a Feshbach resonance technique and kinetic energy by
changing the lattice depth allows us to capture such Hubbard physics in
a broad range of parameter regimes.

The recent achievement of Fermi degeneracy of ultracold
alkaline-earth-metal(-like) atoms such as $^{43}$Ca, $^{87}$Sr, and
$^{173}$Yb potentially provides a new class of strongly correlated
matter. 
The structure of their nuclear spin degrees of freedom allows
the realization of high symmetry groups for the internal degree of
freedom (``spin''). 
For instance, $^{173}$Yb atoms behave as a spin-$5/2$
fermion~\cite{Taie.etal/PRL105.2010,Fukuhara.et.al/PRR98.2007,Cazalilla.Ho.Ueda/New.J.Phys.11.2009}
and $^{87}$Sr as a spin-$9/2$
fermion~\cite{Ye.et.al/Science320.2008,Hermele.Gurarie.Rey/PRL103.2009,Hermele.Gurarie.Rey/PRL107.2011/Erratum,DeSalvo.et.al/PRL105.2010}.
In particular, provided all $s$-wave scattering lengths are
independent of the atomic spin states, the atom cloud as a many-body
system obeys high symmetries. Thus the confinement of
alkaline-earth-metal(-like) atoms in an optical lattice provides
opportunities for the study of the SU($N$) symmetric Hubbard model for
spin-$(N-1)/2$ atoms.
The experimental realization of such SU($N$) Hubbard models has strongly
stimulated the corresponding theoretical
studies~~\cite{Hermele.Gurarie.Rey/PRL103.2009,Hermele.Gurarie.Rey/PRL107.2011/Erratum,Manmana/PRA84.2011,Honerkamp.Hofstetter/PRL92.2004,Cazalilla.Ho.Ueda/New.J.Phys.11.2009,Gorshkov.etal/NatPhys.6.2010,Hazzard.et.al/PRA85.2012,Nonne.etal/PRB81.2010,Nonne.etal/ModPhysLett25.2011,Nonne.etal/PRB84.2011,Azaria.Capponi.Lecheminant/PRA80.2009}.

In condensed matter physics, the SU($N$) symmetric systems
have been introduced as a purely theoretical extension of the strongly
correlated electron systems with SU($2$) spin rotational symmetry, e.g.,
for quantum
magnetism~\cite{Arovas.Auerbach/PRB38.1988,Read.Sachdev/PRL62.1989,Read.Sachdev/PRB42.1990,Harada.Kawashima.Troyer/PRL90.2003,Kawashima.Tanabe/PRL98.2007}
or for the Hubbard model~~\cite{Ruckenstein.Schmitt-Rink/RPRB38.1988} in
the context of high-$T_c$
superconductivity~\cite{Affleck.Marston/PRB37.1988,Marston.Affleck/PRB39.1989,Lavagna.Millis.Lee/PRL58.1987,Kotliar.Liu/PRL61.1988}.
As a theoretical tool to solve such problems, the
slave-particle technique, originally developed for the single impurity
Anderson 
model~\cite{Abrikosov/Physics2,Barnes/JPhysF6.1976,Read.Newns/JPhysC16.1983,Coleman/PRB29.1984},
has been used and applied to the Hubbard
model~\cite{Kotliar.Ruckenstein/PRL57.1989,Li.Woelfle.Hirschfeld/PRB40.1989,Lavagna/PRB41.1990,Kane.Lee.Read/PRB39.1989,Marsiglio.et.al/PRB43.1991}.
More recently, the slave-boson approach has also been used with success
in the cold atom
context~\cite{Sensarma.etal/PRL103.2009,Huber.Ruegg/PRL.102.2009,Tokuno.Demler.Giamarchi/PRA85.2012}.

In this paper, we generalize the slave-boson calculation scheme
introduced in Ref.~\cite{Tokuno.Demler.Giamarchi/PRA85.2012} to
the symmetric $N$-component fermionic atom systems including SU($N$)
symmetry. 
This technique has proven an effective way to compute the dynamics of
strongly interacting systems at a filling of one or less than one
particle per site in the spin-incoherent temperature regime for which
the temperature is lower than the interaction energy, but larger than
the magnetic superexchange one. 
This regime is directly relevant to the current experiments on fermionic
atoms. 
A diagrammatic approach based on the noncrossing approximation (NCA)
with the spin-incoherent assumption is used to estimate self-energies
and compute the spectral functions as functions of temperature, chemical
potential, and component number $N$. 
This allows us in particular to compare the physics for different $N$s.

We also use these techniques to compute the effect of lattice modulation
spectroscopy which has been recently implemented in experiments.
The lattice modulation technique has been originally applied to bosonic
atom systems, in which the absorbed energy is measured as a function of
the modulation frequency~\cite{Stoeferle.et.al/PRL92.2004}.
According to the linear response formalism, the energy absorption rate
in such a weak perturbation regime gives access to the kinetic-energy
correlation
functions~\cite{Iucci.etal/PRA73.2006,Reischl.Schmidt.Uhrig/PRA72.2005,Kollath.etal/PRA74.2006}.
For fermionic atoms, an accurate measurement of the absorption
energy is difficult, and a variant of the probe measuring the
production rate of so-called doublons, which is the number of doubly
occupied sites, induced by the lattice modulation has been
proposed.~\cite{Kollath.etal/PRA74.2006} 
The doublon production rate (DPR) has been shown to be identical to the
energy absorption rate both numerically and
analytically~\cite{Kollath.etal/PRA74.2006,Tokuno.Gimarchi/PRL106.2011,Tokuno.Giamarchi/PRA85.2012}.
This doublon measurement technique has been successfully implemented in
a fermionic atom experiment~\cite{Joerdens.etal/Nature455.2008}.
This allowed more recent experiments to successfully reach the linear
response regime for which the DPR spectrum scales quadratically 
with the modulation amplitude.~\cite{Greif.etal/PRL106.2011}
A direct comparison with equilibrium theory is possible and has been
successfully
done~\cite{Sensarma.etal/PRL103.2009,Tokuno.Demler.Giamarchi/PRA85.2012}. 
So far the lattice modulation experiment has been done with
$^{40}$K~\cite{Greif.etal/PRL106.2011} behaving as a spin-$1/2$ fermion
and $^{173}$Yb~\cite{Taie.etal/NatPhys2012} behaving as a spin-$5/2$
fermion.

The paper is organized as follows.
In Sec.~\ref{sec:model}, we introduce the Hubbard model for
the $N$-component fermions which includes SU($N$) symmetry.
The introduced Hamiltonian is rewritten in a slave-particle representation.
In Sec.~\ref{sec:single-particle-property}, we discuss the single
particle properties based on the Hubbard Hamiltonian using this slave-particle
representation.
Then, using a diagrammatic approach starting from the atomic limit,
self-consistent equations of doublon and holon self-energies are
constructed and solved analytically.
In Sec.~\ref{sec:dpr}, using the spectral functions given in
Sec.~\ref{sec:single-particle-property}, we investigate the DPR spectrum
produced by an amplitude modulation of the optical lattice potential.
Then the analytic form of the DPR spectrum is given.
In addition, we extend the calculation to the trapped case by using the
local density approximation (LDA), and we also compare our results to
the recent experiment with $^{173}$Yb atoms.
Finally, our results are summarized in Sec.~\ref{sec:summary}.
Some technical details on the formulation of the DPR spectrum are
briefly reviewed in Appendix~\ref{DPRformalism}.

\section{Model}\label{sec:model}
\subsection{SU($N$)-symmetric Hubbard model}
In lattice systems with multicomponent particles, multiparticle
occupation states in addition to double occupancy can be defined in
general. However, when the interaction between different
components is strong, such multiple-occupation is at a higher energy
state than double occupancy.
Because we are interested in physics of doublon excitations at a filling
of one or less than one particle per site, such higher occupation states
are way above the main energy scale of interest.
Thus, as an effective model Hamiltonian, we can extend the
Hubbard type two-body interaction to the $N$-component case.
Then interactions between different components, determined by the
$s$-wave scattering lengths, generally take different values depending
on the components:
The interaction parameter is written as $U_{\sigma,\sigma'}$ where
$\sigma$ and $\sigma'$ are the indices characterizing the internal state
of the fermions. 
We consider the special case of a unique interaction parameter,
$U_{\sigma,\sigma'}=U$~\cite{Kitagawa.etal/PRA77.2008,Stellmer.Grimm.Schreck/PRA84.2011};
namely the coupling does not depend on the components.~\footnote{For example, it is known that for $^{173}$Yb atoms, which are spin-$5/2$ fermions, all two-body interactions are the same.}
Then, the interaction term has the same symmetry as the kinetic term,
and the system symmetry turns out to be enlarged to SU($N$) symmetry.

We consider the generalized $N$-component fermionic Hubbard model,
$H=H_{\rm K}+H_{\rm at}$ with
\begin{equation}
 \begin{aligned}
  H_{\rm K}
  &=-J\sum_{\sigma=1}^{N}\sum_{\langle{i,j}\rangle}
  c^{\dagger}_{i,\sigma}c_{j,\sigma},
  \\
  H_{\rm at}
  &=-\mu\sum_{\sigma=1}^{N}\sum_{j}n_{j,\sigma}
  +\sum_{\sigma\ne\sigma'}\sum_{j}
    \frac{U}{2}n_{j,\sigma}n_{j,\sigma'},
 \end{aligned}
 \label{eq:Hubbard_model}
\end{equation}
where $c_{j,\sigma}$ is the annihilation operator of a fermion with the
internal component $\sigma$ at a site $j$, and $n_{j,\sigma}$ is the
number operator. 
The parameters $J$ and $U$, respectively, denote the nearest-neighbor
hopping energy and the on-site interaction between components $\sigma$ and
$\sigma'$.
Throughout this paper, we consider only the repulsive case $U>0$.

In the considered regime of chemical potentials, particle-hole symmetry
always disappears except for the $N=2$ case at $\mu=U/2$. 
This is because for $N>2$ there are $N(N-1)/2$ doublon states while the
hole state is unique.

\subsection{slave particle representation}

The $N$-component fermions have a larger Hilbert space than that of the
two-component case:
While an empty site (holon) is unique, there exist multiple-occupation
states (three- and four-fold occupation, and so on) in addition to
$N(N-1)/2$-species doubly occupied states (doublons) and $N$-species
single occupied spin states (spinons).
The multiple-occupation states are energetically out of shell, since
those states cost an energy higher than $2U$, which is over the energetic
cutoff in our model Hamiltonian.
Thus in our case we describe the single-site state by a holon $\ket{h}$,
$N$-species spinon $\ket{\sigma}$ ($\sigma=1,2,\cdots,N$), and
$N(N-1)/2$-species doublons $\ket{d_{\sigma,\sigma'}}$
($\sigma\ne\sigma'$ and $\sigma,\sigma'=1,\cdots,N$).
In terms of doublon states, the antisymmetrization condition
$\ket{d_{\sigma,\sigma'}}=-\ket{d_{\sigma',\sigma}}$ is imposed.
We hereafter suppose that the spin indices in $\ket{d_{\sigma,\sigma'}}$
are ordered such that $\sigma>\sigma'$.
The single-site original fermionic operators, $c_{\sigma}$ and
$c_{\sigma}^{\dagger}$, are given by
\begin{align}
 c_{\sigma}
 &=\ket{h}\bra{\sigma}
   +\sum_{\sigma'=1}^{\sigma-1}\ket{\sigma'}\bra{d_{\sigma,\sigma'}}
   -\sum_{\sigma'=\sigma+1}^{N}\ket{\sigma'}\bra{d_{\sigma',\sigma}},
 \\
 c_{\sigma}^{\dagger}
 &=\ket{\sigma}\bra{h}
   +\sum_{\sigma'=1}^{\sigma-1}\ket{d_{\sigma',\sigma}}\bra{\sigma'}
   -\sum_{\sigma'=\sigma+1}^{N}\ket{d_{\sigma,\sigma'}}\bra{\sigma'}.
\end{align}
Then the representation no longer gives back the anticommutation
relation $\{c_{\sigma},c_{\sigma}^{\dagger}\}=\delta_{\sigma,\sigma'}$,
but it should be approximately correct as long as $U$ is much larger
than the particle hopping $J$ and temperature, and the filling is less
than unity, which means that all the excitations leave the system in
the proper subpart of the Hilbert space.
We introduce the creation and annihilation operators for a holon,
spinons, and doublons as
\begin{equation}
 \begin{aligned}
  &\ket{\sigma}\bra{h}=b_{\sigma}^{\dagger}h, \\
  &\ket{h}\bra{\sigma}=h^{\dagger}b_{\sigma}, \\
  &\ket{d_{\sigma,\sigma'}}\bra{\sigma''}=d_{\sigma\sigma'}^{\dagger}b_{\sigma''}, \\
  &\ket{\sigma''}\bra{d_{\sigma,\sigma'}}=b_{\sigma''}^{\dagger}d_{\sigma\sigma'},
 \end{aligned}
\end{equation}
and the new vacuum state $\ket{0}$ is defined as
$b_{\sigma}\ket{0}=h\ket{0}=d_{\sigma\sigma'}\ket{0}=0$.

It is easy to extend the above single-site argument to multi-site
problems: All operators become site dependent.
In order to recover the anticommutation relations between the original
fermions at different sites,
$\{c_{i,\sigma},c_{j,\sigma'}^{\dagger}\}=0$ ($i\ne j$), we assume the
following commutation and anticommutation relations:
$\{h_{i},h_{j}^{\dagger}\}=\delta_{i,j}$,
$\{d_{i,\sigma\sigma'},d_{j,\eta\eta'}^{\dagger}\}=\delta_{i,j}\delta_{\sigma,\eta}\delta_{\sigma',\eta'}$,
and
$[b_{i,\sigma},b_{j,\sigma'}^{\dagger}]=\delta_{i,j}\delta_{\sigma,\sigma'}$.
Furthermore, by imposing the following constraint we project onto the
physical subspace the Hilbert space enlarged by introducing the holon,
spinons, and doublons:
\begin{equation}
 h_{j}^{\dagger}h_{j}
  +\sum_{\sigma=1}^{N}b_{j,\sigma}^{\dagger}b_{j,\sigma}
  +\sum_{\sigma>\sigma'}d_{j,\sigma\sigma'}^{\dagger}d_{j,\sigma\sigma'}
 =1,
 \label{eq:slave_particle_constraint}
\end{equation}
which means that the double occupation on the same site by the slave
particles (holon, spinons and doublons) is forbidden.

In summary, the $N$-component fermion in the reduced Hilbert space
where the multiple-occupied states are truncated is described in slave
particle description as follows:
\begin{equation}
 \begin{aligned}
  c_{j,\sigma}
  &=h_{j}^{\dagger}b_{j,\sigma}
  +\sum_{\sigma'=1}^{\sigma-1}b_{j,\sigma'}^{\dagger}d_{j,\sigma\sigma'}
  -\sum_{\sigma'=\sigma+1}^{N}b_{j,\sigma'}^{\dagger}d_{j,\sigma'\sigma},
  \\
  c_{j,\sigma}^{\dagger}
  &=b_{j,\sigma}^{\dagger}h_{j}
  +\sum_{\sigma'=1}^{\sigma-1}d_{j,\sigma\sigma'}^{\dagger}b_{j,\sigma'}
  -\sum_{\sigma'=\sigma+1}^{N}d_{j,\sigma'\sigma}^{\dagger}b_{j,\sigma'}.
 \end{aligned}
 \label{eq:Slave_Particle_Formula}
\end{equation}
Due to the slave-particle
constraint~(\ref{eq:slave_particle_constraint}) this representation
automatically leads to the expected number operator:
\begin{equation}
 n_{j\sigma}
 =b_{j,\sigma}^{\dagger}b_{j,\sigma}
  +\sum_{\sigma'=1}^{\sigma-1}d_{j,\sigma\sigma'}^{\dagger}d_{j,\sigma\sigma'}
  +\sum_{\sigma'=\sigma+1}^{N}d_{j,\sigma'\sigma}^{\dagger}d_{j,\sigma'\sigma}.
\end{equation}
The constraint~(\ref{eq:slave_particle_constraint}) is imposed
by a Lagrange multiplier method.
The Hamiltonian~(\ref{eq:Hubbard_model}) is represented as
\begin{align}
 H_{\mathrm{K}}
 &=-J\sum_{\langle{i,j}\rangle}
   \Biggl[
     \sum_{\sigma=1}^{N}
          \mathcal{F}_{i,j}^{\sigma\sigma}h_{i}h_{j}^{\dagger}
     +\sum_{\sigma_1>\sigma_2}
         \biggl(
           \mathcal{A}_{i,j}^{\sigma_1\sigma_2\dagger}h_{i}d_{j,\sigma_1\sigma_2}
  \nonumber \\
  & \quad
           +\mathcal{F}_{j,i}^{\sigma_2\sigma_2}d_{i,\sigma_1\sigma_2}^{\dagger}d_{j,\sigma_1\sigma_2}
           +\mathrm{H.c.}
         \biggr)
  \nonumber \\
  & \quad
        +\sum_{\sigma_1>\sigma_2>\sigma_3}
         \biggl(
           \mathcal{F}_{j,i}^{\sigma_3\sigma_2}d_{i,\sigma_1\sigma_2}^{\dagger}d_{j,\sigma_1\sigma_3}
  \nonumber \\
  & \quad
           +\mathcal{F}_{j,i}^{\sigma_2\sigma_1}d_{i,\sigma_1\sigma_3}^{\dagger}d_{j,\sigma_2\sigma_3}
           -\mathcal{F}_{j,i}^{\sigma_1\sigma_3}d_{i,\sigma_2\sigma_3}^{\dagger}d_{j,\sigma_1\sigma_2}
  \nonumber \\
  & \quad
           +\mathrm{H.c.}
         \biggr)
     \Biggr],
  \label{eq:slave_particle_kinetic_energy}\\
  H_{\mathrm{at}}
  &=\sum_{j}
    \biggl[
      -\mu-\lambda_{j}
      +\epsilon_{j}^{\mathrm{h}}h_{j}^{\dagger}h_{j}
      +\sum_{\sigma=1}^{N}
       \epsilon_{j}^{\mathrm{b}}b_{j,\sigma}^{\dagger}b_{j,\sigma}
  \nonumber \\
  & \quad
      +\sum_{\sigma>\sigma'}
       \epsilon_{j}^{\mathrm{d}}d_{j,\sigma\sigma'}^{\dagger}d_{j,\sigma\sigma'}
    \biggr],
  \label{eq:slave_particle_atomic_energy}
 \end{align}
where the local potentials for the slave particles have been introduced
as 
\begin{equation}
 \begin{aligned}
  & \epsilon_{j}^{\mathrm{h}}=\mu+\lambda_{j}, \\
  & \epsilon_{j}^{\mathrm{b}}=\lambda_{j}, \\
  & \epsilon_{j}^{\mathrm{d}}=U-\mu+\lambda_{j},
 \end{aligned}
 \label{eq:local_potential}
\end{equation}
for the holon, spinons and doublons, respectively.
$\lambda_j$ is the Lagrange multiplier for the local constraint.
To simplify the form of $H_{\mathrm{K}}$, a spinon hopping operator from
$i$th to $j$th site, $\mathcal{F}_{j,i}^{\sigma_1\sigma_2}$,
and a creation operator of an antisymmetric spinon pair between
nearest-neighbor sites, $\mathcal{A}_{i,j}^{\sigma_1\sigma_2\dagger}$,
have been defined as
\begin{equation}
 \begin{aligned}
  &\mathcal{F}_{j,i}^{\sigma_1\sigma_2}
   =b^{\dagger}_{j\sigma_1}b_{i\sigma_2},
  \\
  &\mathcal{A}_{i,j}^{\sigma_1\sigma_2\dagger}
   =b^{\dagger}_{i,\sigma_1}b^{\dagger}_{j,\sigma_2}
    -b^{\dagger}_{i,\sigma_2}b^{\dagger}_{j,\sigma_1}.
 \end{aligned}
\end{equation}
The spinon pair operator $\mathcal{A}_{i,j}^{\sigma_1\sigma_2\dagger}$
is the extension of an annihilation operator of a singlet spin
configuration for the two-component case to a generic $N$-component
case.

\section{Single doublon and holon propagator}\label{sec:single-particle-property}

We calculate the single-particle propagator of a holon and a doublon
based on the 
Hamiltonian~(\ref{eq:slave_particle_kinetic_energy}) and
(\ref{eq:slave_particle_atomic_energy}) in the slave-particle
representation.

\subsection{Atomic limit}
Let us start with the atomic limit where $J/U=0$.
Since the atomic Hamiltonian $H_{\mathrm{at}}$ is quadratic in the
slave-particle representation, the atomic propagators of the slave
particles at a $j$th site are immediately given as 
\begin{align}
 &\mathcal{G}^{\mathrm{h}}_{\mathrm{at}}(\bm{r}_j,i\nu_n)
  =\frac{1}{i\nu_n-\epsilon_{j}^{\mathrm{h}}/\hbar},
 \\
 &\mathcal{G}^{\mathrm{b}}_{\mathrm{at}}(\bm{r}_j,i\omega_n)
  =\frac{1}{i\omega_n-\epsilon_{j}^{\mathrm{b}}/\hbar},
 \\
 &\mathcal{G}^{\mathrm{d}}_{\mathrm{at}}(\bm{r}_j,i\nu_n)
  =\frac{1}{i\nu_n-\epsilon_{j}^{\mathrm{d}}/\hbar},
\end{align}
for a holon, spinons, and doublons, respectively.
The atomic propagators of spinons and doublons have the same form
regardless of their species.
$\omega_n$ and $\nu_n$ denote the Matsubara frequency
for bosons and for fermions, respectively.

\subsection{Mean-field assumption of spin-incoherence}
We proceed with the finite- but small-$J$ case by making a
mean-field approximation.
In general, due to the effect of the hopping Hamiltonian, the system
becomes coherent and exhibits a long-range magnetic order.
In order to see such phases, the system should reach a temperature
region lower than the magnetic and charge hopping energy scales.
However, in the spin-incoherent Mott physics case of interest in the
present paper, both spin and charge coherence are expected to be
suppressed due to thermal effects.
This is a common feature of the atomic limit, but to reproduce the
finite bandwidth in terms of single doublon and holon spectra it is
necessary to take into account the effect of the kinetic energy
$H_{\mathrm{K}}$.
As a simple way to describe the spin-incoherent regime, we use the
assumption, which is valid for
$J\ll k_{B}T\ll U$,~\footnote{At a filling of one particle per site, the
system for $k_{B}T\ll U$ would become a Mott insulator. The charge
coherence should then be suppressed even if $k_{B}T<J$. Thus, the
spin-incoherent temperature regime at unity filling is $J^{2}/U\ll
k_{B}T\ll U$.} that the spinon propagation is well described by the
atomic one:
\begin{equation}
 \mathcal{G}^{\mathrm{b}}_{\sigma}(\bm{r}_{j}-\bm{r}_{j'},i\omega_n)
  \rightarrow
  \delta_{j,j'}
  \mathcal{G}^{\mathrm{b}}_{\mathrm{at}}(\bm{r}_j,i\omega_n).
  \label{eq:Spinon_propagator_replacement}
\end{equation}
Note that the atomic propagator does not have translational symmetry in
general. Indeed, $\mathcal{G}_{\mathrm{b}}(\bm{r}_j,i\omega_n)$ includes the
local potential coming from the Lagrange multiplier $\lambda_j$, which
is potentially site dependent.
The mean-field treatment of $\lambda_j$ is required to recover the
translation-invariant paramagnetic background in the above framework.
Thus we replace the Lagrange multiplier by the homogeneous one:
\begin{equation}
 \lambda_{j} \rightarrow \lambda.
 \label{eq:MFAsumption}
\end{equation}
Then, the local potentials~(\ref{eq:local_potential}) also become
homogeneous by definition:
$\epsilon^{\mathrm{h}}_{j}\rightarrow\epsilon^{\mathrm{h}}$,
$\epsilon^{\mathrm{d}}_{j}\rightarrow\epsilon^{\mathrm{d}}$,
and $\epsilon^{\mathrm{b}}_{j}\rightarrow\epsilon^{\mathrm{b}}$.
The mean-field $\lambda$ is determined by
Eq.~(\ref{eq:slave_particle_constraint}) averaged in the atomic limit,
\begin{equation}
 f(\epsilon^{\mathrm{h}})
 +\frac{N(N-1)}{2}f(\epsilon^{\mathrm{d}})
 +Nb(\epsilon^{\mathrm{b}})
 =1,
 \label{eq:MFeq}
\end{equation}
where $f(x)=\frac{1}{e^{x/k_{B}T}+1}$ and $b(x)=\frac{1}{e^{x/k_{B}T}-1}$
are, respectively, the Fermi and Bose distribution functions.
Equation~(\ref{eq:MFeq}) is a saddle-point equation which minimizes the
atomic-limit free energy.
This self-consistent equation can be analytically solved for
$k_{B}T\ll U$:
\begin{align}
 e^{\lambda/k_B T}
 &=\frac{N+1}{2}
   -\frac{(N+1)(N-2)}{4}e^{-(U-\mu)/k_B T}
 \nonumber \\
 & \quad
   +\biggl[
      \left(\frac{N+1}{2}\right)^2
         +Ne^{-\mu/k_B T}
 \nonumber \\
 & \quad
         -\frac{N^3+2N^2-9N-6}{4}e^{-(U-\mu)/k_B T}
 \nonumber \\
 & \quad
         +\left\{\frac{(N+1)(N-2)}{4}e^{-(U-\mu)/k_B T}\right\}^2
    \biggr]^{1/2}.
 \label{eq:MFlambda}
\end{align}
For $N=2$, it gives back the analytic form given in
Ref.~\cite{Tokuno.Demler.Giamarchi/PRA85.2012}.
As long as $J\ll k_{B}T\ll U$,
the atomic limit provides the suitable physics.
Therefore in the spin-incoherent region the above mean-field theory
works well even if the hopping $J$ is finite.

Within the mean-field assumption~(\ref{eq:MFAsumption}), the
atomic propagator of the spinons also becomes site independent:
$\mathcal{G}^{\mathrm{b}}_{\mathrm{at}}(\bm{r}_{j},i\omega_n)\rightarrow\bar{\mathcal{G}}^{\mathrm{b}}_{\mathrm{at}}(i\omega_n)$.
We can thus use the following form for the spinon propagator:
\begin{equation}
 \mathcal{G}^{\mathrm{b}}_{\sigma}(\bm{k},i\omega_n)
  \rightarrow
  \bar{\mathcal{G}}^{\mathrm{b}}_{\mathrm{at}}(i\omega_n)
  \equiv
  \frac{1}{i\omega_n-\epsilon^{\mathrm{b}}/\hbar}.
  \label{eq:Spinon_propagator_replacement2}
\end{equation}
At variance with usual mean-field theory, we include here the dynamical
fluctuations.
The local spin dynamics coming from the thermal fluctuation is thus
retained in this approximation. 

\subsection{Non-crossing approximation}
\begin{figure}[tbp]
 \begin{center}
 \includegraphics[scale=1.]{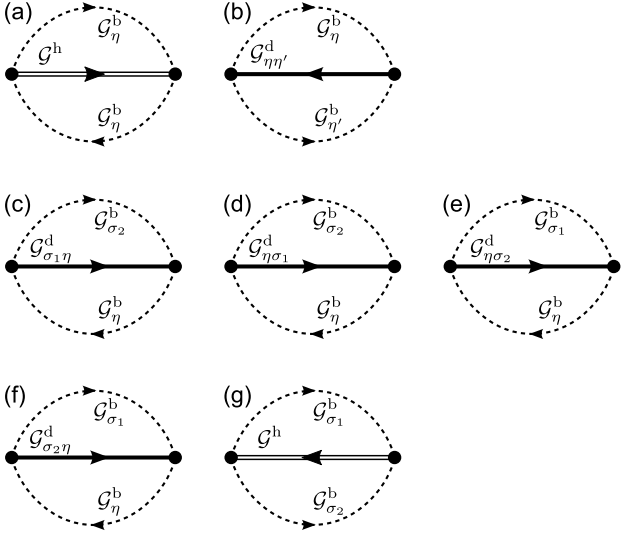}
  \caption{The possible NCA self-energy diagrams for a holon 
  [(a) and (b)] and for doublons [(c)--(g)].
  The solid, double-solid, and dashed lines denote the full propagators
  of the doublon, the holon, and the spinon, respectively.
  The indices $\eta$ and $\eta'$ appearing in the doublon and spinon
  propagators are dummy spin ones, in which the summation
  over possible spin values is taken.
  The holon self-energies $\Sigma^{\mathrm{h}(1)}$ and
  $\Sigma^{\mathrm{h}(2)}$ correspond to (a) and (b).
  One of the two parts of the doublon self-energy
  $\Sigma^{\mathrm{d}(1)}_{\sigma_1\sigma_2}$ includes
  (c)--(f), and the other $\Sigma^{\mathrm{d}(2)}_{\sigma_1\sigma2}$
  is illustrated by (g).
  In the simple NCA idea, each propagator should be dealt with as the
  full ones, but in the approximation shown in this paper, the spinon
  propagators are replaced by the bare, namely, atomic-limit, ones.
  In addition, diagrams (b) and (g) are off-shell, because they must
  not be relevant in the energy regime $\sim U$ since $U$ is very
  large.
  Thus in this paper diagrams~(a) and~(c)--(f) are taken into
  consideration.}
  \label{fig:NCA_Diagrams}
 \end{center}
\end{figure}

Let us consider the full doublon and holon propagators, based on the
atomic-limit mean-field.
To take into account $H_{\mathrm{K}}$ at a filling of one or less than
one particle per site, we use the NCA~\cite{Kane.Lee.Read/PRB39.1989}.
This method gives a result similar to that from the retraceable path
approximation by Brinkman and Rice~\cite{Brinkman.Rice/PRB2.1970}, and
is reasonably tractable and accurate to describe the physics of single
hole motion in a MI background.
In addition, the NCA allows for the control of the chemical potential and
temperature, which means that one can extend the calculation to an
inhomogeneous case by the LDA.
The NCA diagrams contributing to the self-energy of a doublon and holon are
shown in Figs.~\ref{fig:NCA_Diagrams}(c)--~\ref{fig:NCA_Diagrams}(g).
The self-energy for doublons is given by two types of diagrams
$\Sigma^{\mathrm{d}(1)}_{\sigma_1\sigma_2}$ and
$\Sigma^{\mathrm{d}(2)}_{\sigma_1\sigma_2}$:
$\Sigma^{\mathrm{d}(1)}_{\sigma_1\sigma_2}$ comes from the
scattering among doublons and spinons,
and $\Sigma^{\mathrm{d}(2)}_{\sigma_1\sigma_2}$ involves the process in
which a holon is produced and absorbed.
The holon self-energy can be constructed in a similar way.
The two parts of the self-energy diagrams, $\Sigma^{\mathrm{h}(1)}$ and
$\Sigma^{\mathrm{h}(2)}$, are illustrated in
Figs.~\ref{fig:NCA_Diagrams}(a) and~\ref{fig:NCA_Diagrams}(b).

\begin{figure}[tbp]
 \begin{center}
  \includegraphics[scale=1.]{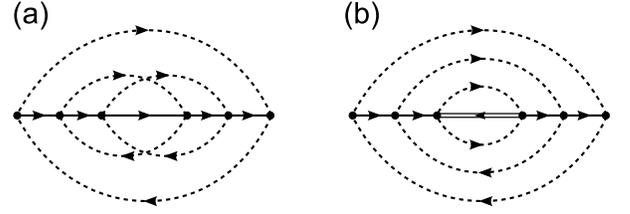}
  \caption{Two examples of diagrams which are not included in
  our NCA: (a) an example of crossing diagrams and (b) one of the
  off-shell diagrams which contains intermediate processes at higher
  energies than the main energy scale, i.e., creation of doublons in the
  holon propagation.
  The solid, double-solid, and dashed lines denote the propagators of
  the holon, the doublon, and the spinon, respectively.
  Note that the doublon and holon propagators shown here mean a bare
  propagator, while all lines denote full propagators in
  Fig.~\ref{fig:NCA_Diagrams}.}
  \label{fig:not-included_diagrams}
 \end{center}
\end{figure}

In principle, the self-energies are determined by solving a set of
self-consistent equations for the doublons and the holon.
The self-energies $\Sigma^{\mathrm{h}(2)}$ and
$\Sigma^{\mathrm{d}(2)}_{\sigma_1\sigma_2}$ link the doublon and holon
propagators as seen in Figs.~\ref{fig:NCA_Diagrams}(b)
and~\ref{fig:NCA_Diagrams}(g).
However, $\Sigma^{\mathrm{h}(2)}$ and
$\Sigma^{\mathrm{d}(2)}_{\sigma_1\sigma_2}$ can be neglected in the present
case, as demonstrated below.
In the strongly interacting case, the two diagrams
Fig.~\ref{fig:NCA_Diagrams}~(b) and~(g) are off-shell because the
intermediate processes creating an additional holon and doublon cost an
additional energy $\sim U$.
Thus, as long as we focus on the physics around the energy scale $\sim U$,
the contribution of such diagrams should be negligible.
In particular, this approximation is expected to be very good when
the system is in a MI state at a filing of one particle per site.
In Fig.~\ref{fig:not-included_diagrams}, we show examples of the
diagrams which are neglected in our NCA calculation.

Consequently, the self-consistent equations of the self-energy of a
holon and a doublon are decoupled and given as
\begin{align}
 \Sigma^{\mathrm{h}}(\bm{k},i\nu_n)
 &=\left(\frac{W_{\mathrm{h}}}{2\hbar}\right)^2
   \frac{1}{\mathcal{V}}\sum_{\bm{q}}
   \mathcal{G}^{\mathrm{h}}(\bm{q},i\nu_n),
 \label{eq:Holon_self-energy}\\
 \Sigma_{\sigma_1\sigma_2}^{\mathrm{d}}(\bm{k},i\nu_n)
 &=\frac{\left(W_{\mathrm{d}}/2\hbar\right)^2 }{2(N-1)}
  \frac{1}{\mathcal{V}}\sum_{\bm{q}}
  \biggl[
    \sum_{\eta=\sigma_2+1}^{\sigma_1-1}\mathcal{G}_{\sigma_1\eta}^{\mathrm{d}}(\bm{q},i\nu_n)
 \nonumber \\
 & \quad
    +\sum_{\eta=\sigma_1+1}^{N}\mathcal{G}_{\eta\sigma_2}^{\mathrm{d}}(\bm{q},i\nu_n)
    +\sum_{\eta=1}^{\sigma_2-1}\mathcal{G}_{\sigma_2\eta}^{\mathrm{d}}(\bm{q},i\nu_n)
 \nonumber \\
 & \quad
    +N\mathcal{G}_{\sigma_1\sigma_2}^{\mathrm{d}}(\bm{q},i\nu_n)
  \biggr],
 \label{eq:Doublon_self-energy}
\end{align}
where $\mathcal{V}$ is the total number of lattice sites.
We have also introduced:
\begin{equation}
 \begin{aligned}
  W_{\mathrm{h}}
  &=\sqrt{4Nzb(\epsilon^{\mathrm{b}})[b(\epsilon^{\mathrm{b}})+1]}J,
  \\
  W_{\mathrm{d}}
  &=\sqrt{8(N-1)zb(\epsilon^{\mathrm{b}})[b(\epsilon^{\mathrm{b}})+1]}J,
 \end{aligned}
 \label{eq:bandwidth}
\end{equation}
which correspond to the half bandwidths of the lower and upper Hubbard
bands, respectively, as seen below.
Because spontaneous breaking of the SU($N$) symmetry is unlikely in the
spin-incoherent temperature region, the single-particle property of the
doublons is independent of the doublon species:
\begin{equation}
 \begin{aligned}
  &\mathcal{G}^{\mathrm{d}}_{\sigma_1\sigma_2}(\bm{k},i\nu_n)
   \equiv\mathcal{G}^{\mathrm{d}}(\bm{k},i\nu_n),
  \\
  &\Sigma^{\mathrm{d}}_{\sigma_1\sigma_2}(\bm{k},i\nu_n)
   \equiv\Sigma^{\mathrm{d}}(\bm{k},i\nu_n).
 \end{aligned}
 \label{eq:SU(N)_assumption}
\end{equation}
Then Eq.~(\ref{eq:Doublon_self-energy}) is simplified more as
\begin{equation}
  \Sigma^{\mathrm{d}}(\bm{k},i\nu_n)
 =\left(\frac{W_{\mathrm{d}}}{2\hbar}\right)^2
  \frac{1}{\mathcal{V}}\sum_{\bm{q}}
   \mathcal{G}^{\mathrm{d}}(\bm{q},i\nu_n),
 \label{eq:Doublon_self-energy2}
\end{equation}
which has a form similar to that of Eq.~(\ref{eq:Holon_self-energy}).
The form of the self-consistent equations (\ref{eq:Holon_self-energy})
and (\ref{eq:Doublon_self-energy2}) leads to the momentum independence
of the self-energies, and thus of the propagators.
Using the Dyson equation, the self-consistent equation is analytically
solved, and the resulting propagators are
\begin{equation}
 \mathcal{G}^{\mathrm{h/d}}(i\nu_n)
 =\frac{2}{i\nu_n-\frac{\epsilon^{\mathrm{h/d}}}{\hbar}+\sqrt{\left(i\nu_n-\frac{\epsilon^{\mathrm{h/d}}}{\hbar}\right)^2-\left(\frac{W_{\mathrm{h/d}}}{\hbar}\right)^2}}.
\end{equation}
Finally, the spectral functions are obtained~\footnote{In the present case, the spectral function is a local quantity. Thus it corresponds to the density of state.} via analytic continuation and are:
\begin{equation}
 \mathcal{A}^{\mathrm{h/d}}(\omega)
 =\frac{W_{\mathrm{h/d}}}{4\hbar}
  \sqrt{1-\left(\frac{\hbar\omega-\epsilon^{\mathrm{h/d}}}{W_{\mathrm{h/d}}}\right)^2}.
 \label{eq:spectral_funcsions}
\end{equation}
The spectral function shows the same semicircular behavior as for a
single hole in a half-filled $t$-$J$ model discussed in
Ref.~\cite{Kane.Lee.Read/PRB39.1989}. 
Note that in that case the slave bosons would be
condensed as a consequence of the long-range antiferromagnetic order.
In addition, this result is similar to that of the retraceable path
approximation~\cite{Brinkman.Rice/PRB2.1970}.

\begin{figure}[tbp]
 \begin{center}
  \includegraphics[scale=1.]{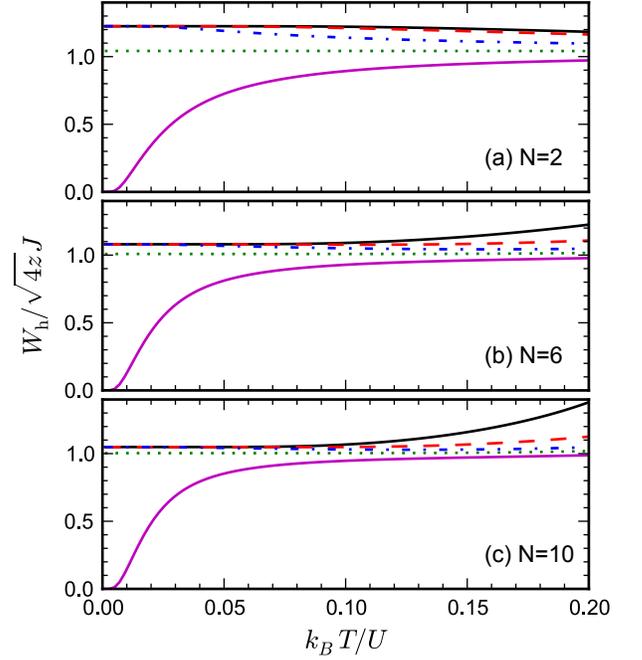}
  \caption{(Color online)
  The holon bandwidth $W_{\mathrm{h}}/\sqrt{4z}J$ for different
  chemical potentials as a function of temperature and chemical
  potential:
  (a) for $N=2$, (b) for $N=6$, and (c) for $N=10$, respectively.
  The lines from upper to lower denote $\mu/U=0.5$, $0.3$, $0.1$, $0.0$,
  and $-0.1$, respectively.
  }
  \label{fig:band-width}
 \end{center}
\end{figure}

The bandwidth~(\ref{eq:bandwidth}) as a function of temperature and
chemical potential is shown in Fig.~\ref{fig:band-width}.
$W_{\mathrm{h}}/W_{\mathrm{d}}$ depends only on $N$:
$W_{\mathrm{d}}/W_{\mathrm{h}}=\sqrt{2(N-1)/N}$, which monotonically
increases and asymptotically reaches $\sqrt{2}$ as $N$ goes up.
$W_{\mathrm{d}}$ is larger than $W_{\mathrm{h}}$ for $N>2$.
As in Ref.~\cite{Tokuno.Demler.Giamarchi/PRA85.2012}, in the
SU($2$) case, the shape of the two bands is the same.
Figure~\ref{fig:band-width} shows that the temperature dependence is
different depending on whether or not $N=2$.
While the bandwidth for $N>2$ monotonically increases with temperature,
it decreases for $N=2$ and $\mu/U>0$.

\begin{figure*}[tbp]
 \begin{center}
  \includegraphics[scale=.24]{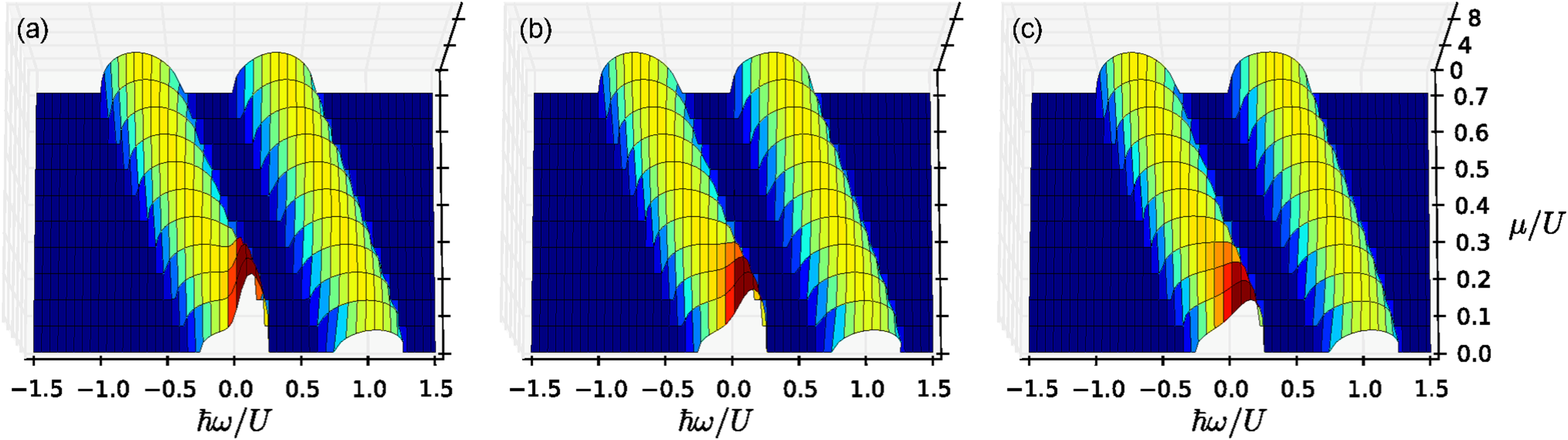}
  \caption{(Color online)
  The spectral functions of the fermionic atoms for $N=2$ as a
  function of the chemical potential for different temperatures:
  (a) $k_{B}T<J$ ($k_{B}T/U=0.025$, $J/U=0.05$),
  (b) $k_{B}T=J$ ($k_{B}T/U=J/U=0.05$),
  and (c) $k_{B}T>J$ ($k_{B}T/U=0.075$, $J/U=0.05$).
  The chemical potential runs from $\mu/U=0$ to $0.7$.
  }
  \label{fig:spectralfunctionN=2}
 \end{center}
 \begin{center}
  \includegraphics[scale=.24]{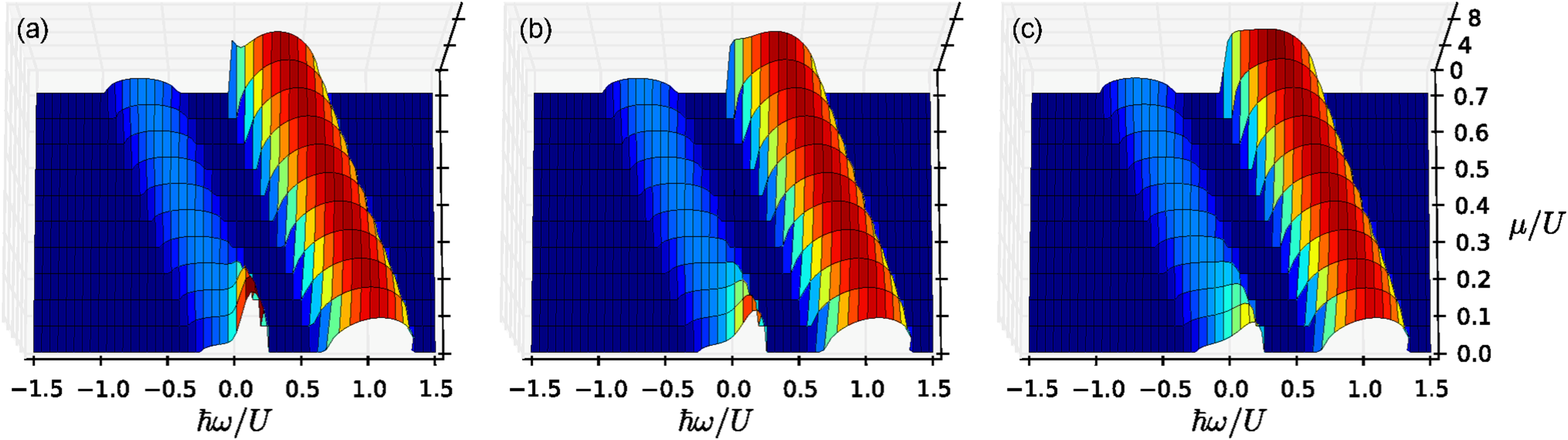}
  \caption{(Color online)
  The spectral functions of the fermionic atoms for $N=6$ as a
  function of the chemical potential for different temperatures:
  (a) $k_{B}T<J$ ($k_{B}T/U=0.025$, $J/U=0.05$),
  (b) $k_{B}T=J$ ($k_{B}T/U=J/U=0.05$),
  and (c) $k_{B}T>J$ ($k_{B}T/U=0.075$, $J/U=0.05$).
  The chemical potential runs from $\mu/U=0$ to $0.7$.
  }
  \label{fig:spectralfunctionN=6}
 \end{center}
 \begin{center}
  \includegraphics[scale=.24]{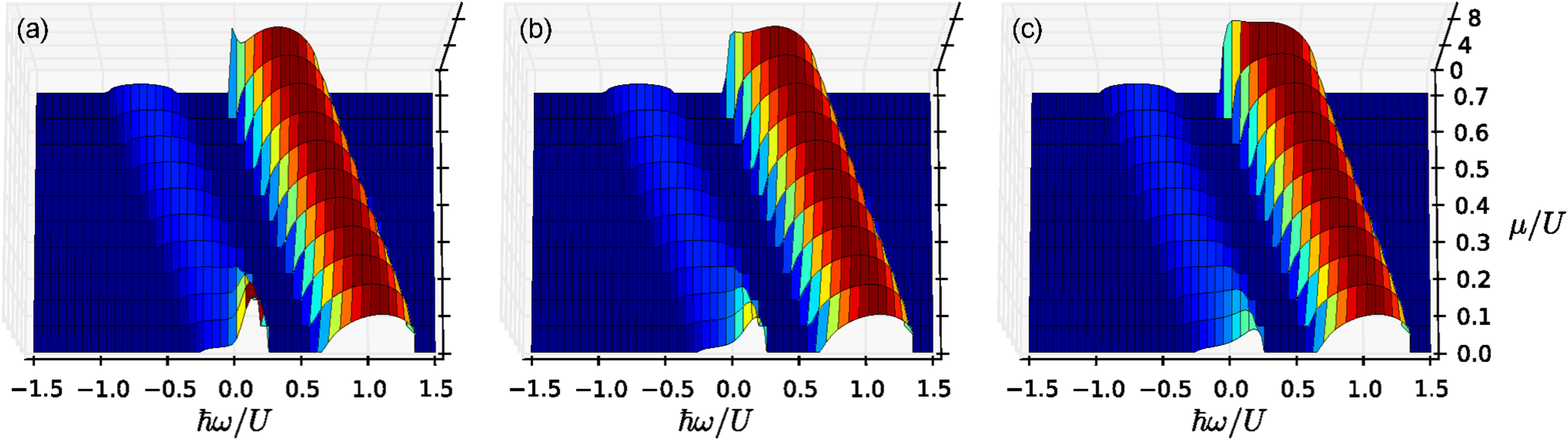}
  \caption{(Color online)
  The spectral functions of the fermionic atoms for $N=10$ as a
  function of the chemical potential for different temperatures:
  (a) $k_{B}T<J$ ($k_{B}T/U=0.025$, $J/U=0.05$),
  (b) $k_{B}T=J$ ($k_{B}T/U=J/U=0.05$),
  and (c) $k_{B}T>J$ ($k_{B}T/U=0.075$, $J/U=0.05$).
  The chemical potential runs from $\mu/U=0$ to $0.7$.
  }
  \label{fig:spectralfunctionN=10}
 \end{center}
\end{figure*}

Let us look at the single-particle properties of the original fermions.
Using the slave-particle
representation~(\ref{eq:Slave_Particle_Formula}), the Matsubara Green
function of the original fermion is expressed as
\begin{align}
 \mathcal{G}_{\sigma,\sigma'}(\bm{r}_{j,j'},\tau)
 &=-\langle{T_{\tau}h^{\dagger}_{j}(\tau)h_{j}(0)}\rangle
   \langle{T_{\tau}b_{j,\sigma}(\tau)b_{j',\sigma'}^{\dagger}(0)}\rangle
 \nonumber \\
 &\quad
  -\sum_{\eta=1}^{\sigma-1}\sum_{\eta'=1}^{\sigma'-1}
   \langle{T_{\tau}d_{j,\sigma\eta}(\tau)d^{\dagger}_{j',\sigma'\eta'}(0)}\rangle
 \nonumber \\
 &\quad \times
   \langle{T_{\tau}b_{j,\eta}^{\dagger}(\tau)b_{j',\eta'}(0)}\rangle
 \nonumber \\
 &\quad
  -\sum_{\eta=\sigma+1}^{N}\sum_{\eta'=\sigma'+1}^{N}
   \langle{T_{\tau}d_{j,\eta\sigma}(\tau)d_{j',\eta'\sigma'}^{\dagger}(0)}\rangle
 \nonumber \\
 &\quad \times
   \langle{T_{\tau}b_{j,\eta}^{\dagger}(\tau)b_{j',\eta'}(0)}\rangle
 \nonumber \\
 &\quad
  +\sum_{\eta=1}^{\sigma-1}\sum_{\eta'=\sigma'+1}^{N}
   \langle{T_{\tau}d_{j,\sigma\eta}(\tau)d_{j',\eta'\sigma}^{\dagger}(0)}\rangle
 \nonumber \\
 &\quad \times
   \langle{T_{\tau}b_{j,\eta}^{\dagger}(\tau)b_{j',\eta'}(0)}\rangle
 \nonumber \\
 &\quad
  +\sum_{\eta=\sigma+1}^{N}\sum_{\eta'=1}^{\sigma'-1}
   \langle{T_{\tau}d_{j,\eta\sigma}(\tau)d_{j',\sigma'\eta'}^{\dagger}(0)}\rangle
 \nonumber \\
 &\quad \times
   \langle{T_{\tau}b_{j,\eta}^{\dagger}(\tau)b_{j',\eta'}(0)}\rangle,
\end{align}
where $\bm{r}_{j,j'}=\bm{r}_{j}-\bm{r}_{j'}$, and $T_{\tau}$ denotes the
imaginary time order.
By applying the mean-field
assumption~(\ref{eq:Spinon_propagator_replacement2}) and replacing
the atomic propagator in terms of the spinon, the propagator of the
original fermion,
$\mathcal{G}_{\sigma,\sigma'}(\bm{r}_{j,j'},\tau)$, is found
to be nonzero only for $\sigma=\sigma'$ and $\bm{r}_j=\bm{r}_{j'}$.
We thus take
$\mathcal{G}_{\sigma,\sigma'}(\bm{r}_{j,j'},\tau)=\delta_{j,j'}\delta_{\sigma,\sigma'}\mathcal{G}_{\sigma}(\tau)$.
In addition, the assumption of SU($N$)
symmetry~(\ref{eq:SU(N)_assumption}) simplifies more the form of the
original fermion propagator.
The Fourier transform of the propagator is thus written as
\begin{align}
 \mathcal{G}_{\sigma}(i\nu_n)
 &=\frac{1}{\hbar\beta}\sum_{\omega_{l}}
    \bar{\mathcal{G}}^{\mathrm{b}}_{\mathrm{at}}(i\omega_l)
    \biggl[
     \mathcal{G}^{\mathrm{h}}(i\omega_{l}-i\nu_n)
 \nonumber \\
 &\quad
     -(N-1)\mathcal{G}^{\mathrm{d}}(i\omega_{l}+i\nu_n)
    \biggr],
\end{align}
and by analytic continuation and Lehmann representation, one can obtain
the spectral function as
\begin{align}
 \mathcal{A}_{\sigma}(\bm{k},\omega)
 &=\biggl[
     f(\epsilon^{\mathrm{b}}-\hbar\omega)+b(\epsilon^{\mathrm{b}})
   \biggr]
   \mathcal{A}_{\mathrm{h}}(\epsilon^{\mathrm{b}}/\hbar-\omega)
 \nonumber \\
 &\quad
  +\biggl[
     f(\epsilon^{\mathrm{b}}+\hbar\omega)+b(\epsilon^{\mathrm{b}})
   \biggr]
   (N-1)\mathcal{A}_{\mathrm{d}}(\epsilon^{\mathrm{b}}/\hbar+\omega).
\end{align}
The spectral functions for $N=2$, $6$, and $10$ are, respectively, shown
in Figs.~\ref{fig:spectralfunctionN=2},~\ref{fig:spectralfunctionN=6},
and~\ref{fig:spectralfunctionN=10}.
This analytic form implies, as mentioned above, that the holon and
doublon spectra form the lower and upper Hubbard bands, respectively.
Since the centers of the lower and upper bands are located,
respectively, at
$\omega=(\epsilon^{\mathrm{d}}-\epsilon^{\mathrm{b}})/\hbar=(U-\mu)/\hbar$
and $-(\epsilon^{\mathrm{h}}-\epsilon^{\mathrm{b}})/\hbar=\mu/\hbar$,
the band gap is $U-(W_{\mathrm{h}}+W_{\mathrm{d}})$.
For $N>2$, the upper and lower Hubbard bands are always asymmetric
because of the absence of particle-hole symmetry.
In addition, the weight of the doublon band is larger than that of the
holon, because the possible number of doublon states increases with the
species of the doublons.

\section{Doublon production rate}\label{sec:dpr}
\begin{figure}[tbp]
 \begin{center}
  \includegraphics[scale=1.]{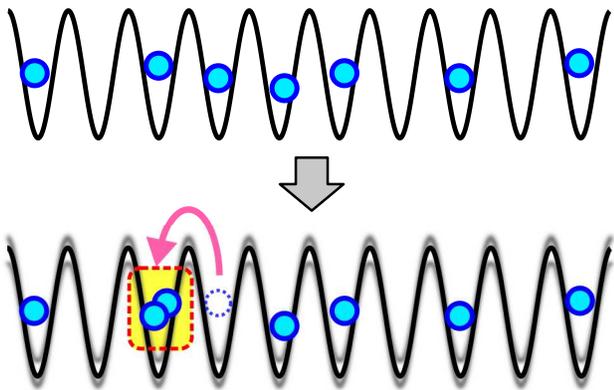}
  \caption{(Color online)
  A sketch of optical lattice modulation and double occupancy.
  Due to the modulation perturbation, the system is excited, and doubly
  occupied sites are created.
  In experiments, the number of formed atom pairs is measured as a
  function of the modulation time duration, and the production rate is
  estimated.}
  \label{fig:lattice-modulation}
 \end{center}
\end{figure}

Using the obtained spectral functions~(\ref{eq:spectral_funcsions}),
we calculate the DPR spectrum of the optical lattice modulation.
In the spectroscopy, the amplitude of an optical lattice in which the atom cloud
is confined is modulated, and the created double occupancy is measured as
shown in Fig.~\ref{fig:lattice-modulation}.
As shown in the Appendix, the DPR per site can be obtained
from a second-order calculation~\footnote{Here the DPR is defined as the number of doubly occupied sites. In Ref.~\cite{Greif.etal/PRL106.2011}, it is defined as the number of atoms forming doublons.} as
\begin{equation}
 P_{D}(\omega)
 =-\frac{(\delta{F})^2}{2\hbar\mathcal{V}U}\omega
   \mathrm{Im}\chi_{\mathrm{K}}^{\mathrm{R}}(\omega),
 \label{eq:DPRformula}
\end{equation}
where $\chi_{\mathrm{K}}(\omega)$ is the Fourier transform of the
retarded correlation function of the kinetic energy
$\chi_{\mathrm{K}}^{\mathrm{R}}(t)=-i\theta(t)\langle{[T_{t}H_{\mathrm{K}}(t),H_{\mathrm{K}}(0)]}\rangle$,
and $\delta{F}$ is the modulation parameter in the lattice model,
given by $\delta{F}=[dJ/dV-dU/dV]\delta{V}$, where $\delta{V}$ is the
amplitude of the optical lattice modulation.

In order to derive the DPR spectrum formula~(\ref{eq:DPRformula}), the
system is assumed to be homogeneous, so the trap is not included in the Hamiltonian.
It is possible to extend this formulation to an inhomogeneous
case~\cite{Tokuno.Giamarchi/PRA85.2012}.
Then the corresponding response function is replaced by
\begin{equation}
 (\delta{F})^2\chi^{\mathrm{R}}_{\mathrm{K}}(\omega)
 \rightarrow
 -i\int_{0}^{\infty}\!\! dt\,
 \langle{[S(t),S(0)]}\rangle.
\label{eq:generalized-DPR-response-function}
\end{equation}
The operator $S$ is defined as
$S=(\delta{F})H_{\mathrm{K}}-(\delta{U})H_{\mathrm{p}}$ where
$\delta{U}=(dU/dV)\delta{V}$ and $H_{\mathrm{p}}$ is the trap potential
term of the Hamiltonian.
The retarded correlation function is computed using the Hamiltonian
$H+H_{\mathrm{p}}$.
The above formula can be used directly in situations for which a direct
computation of the correlation function in the presence of the trap
potential can be implemented by use of numerics such as Monte Carlo
simulations and density-matrix renormalization group approaches.
However, in general it is not easy to directly deal with the effect of
inhomogeneity, and thus we use the LDA to obtain a tractable
approximation of (\ref{eq:generalized-DPR-response-function}).
In the LDA framework,
formula~(\ref{eq:generalized-DPR-response-function}) would be identical
to the one for the homogeneous case~(\ref{eq:DPRformula}).
In what follows, we use Eq.~(\ref{eq:DPRformula}) to calculate the DPR
spectrum in the same manner as discussed in the previous
paper~\cite{Tokuno.Demler.Giamarchi/PRA85.2012} in which the
inhomogeneity effect of the trap is taken into account by the LDA, and
the obtained result shows good agreement with the experimental
data~\cite{Greif.etal/PRL106.2011}.

The DPR is given by the two-particle correlation function, which
includes vertex corrections.
Here we are in the strongly interacting regime ($J/U\ll1$), and we
ignore the vertex correction as a simple approximation.~\footnote{Vertex
corrections are usually needed to preserve symmetries which are broken in
mean-field calculations. However, in this case there is no spontaneous
symmetry breaking in our mean-field calculation. Thus, the vertex
correction would only give a quantitative correction. In the
strongly interacting case this correction will be small.}

We now compute the retarded correlation function
$\chi_{\mathrm{K}}^{\mathrm{R}}(\omega)$ for fillings of
one or less than one particle per site in the spin-incoherent
intermediate temperature regime.
We start with the corresponding imaginary time correlation
function
$\chi_{\mathrm{K}}(\tau)=-\langle{T_{\tau}H_{\mathrm{K}}(\tau)H_{\mathrm{K}}(0)}\rangle$.
In the same way as in the calculation of the spectral function of the
original fermions, the analytic continuation of the time-ordered
correlation function in imaginary time leads to the retarded correlation
function in real time:
$\chi_{\mathrm{K}}^{\mathrm{R}}(\omega)=\tilde{\chi}_{\mathrm{K}}(i\omega_n\rightarrow\omega+i0^+)$,
where $\tilde{\chi}_{\mathrm{K}}(i\omega_n)$ is a Fourier transform of
$\chi_{\mathrm{K}}(\tau)$.
Contrarily to the case of numerical evaluations of correlation functions
in imaginary time, for which there is no straightforward way to perform
the analytic continuation, here we use our analytic form to do so.
This is definitely one of the advantages of the technique used in the
present paper, when computing frequency- or time-dependent
correlations.

The result for $N=2$ is in extremely good agreement with the
experiment of Ref.~\cite{Greif.etal/PRL106.2011}, as discussed in
Ref.~\cite{Tokuno.Demler.Giamarchi/PRA85.2012}.
As we detail below, a similar analytic calculation can be also done for
the case of $N$-component systems, and this allows for a direct
comparison to experiments, via the LDA for a trapped system.

\begin{figure}[tbp]
 \begin{center}
  \includegraphics[scale=1.]{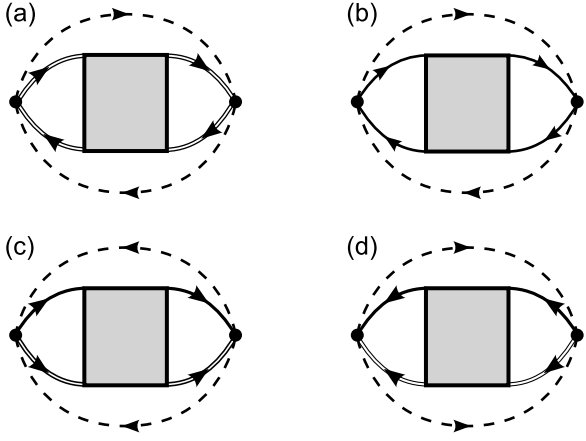}
  \caption{
  Diagrams corresponding to the kinetic-energy correlation functions:
  (a) $\chi_{\mathrm{K}}^{\mathrm{h}}(\tau)$,
  (b) $\chi_{\mathrm{K}}^{\mathrm{d}}(\tau )$,
  and (c) $\chi_{\mathrm{K}}^{\mathrm{hd}}(\tau)$.
  The solid, double-solid, and dashed lines, respectively, denote the
  propagators of the doublon, the holon, and the spinon.
  The shaded rectangle includes a vertex correction which is not
  considered in this paper.
  }
  \label{fig:chi-diagrams}
 \end{center}
\end{figure}

The slave-particle representation is useful to clarify the physical
meaning of the correlation function $\chi_{\mathrm{K}}(\tau)$.
By applying the spin-incoherent
assumption~(\ref{eq:Spinon_propagator_replacement2}), the correlation
$\chi_{\mathrm{K}}(\tau)$ can be written, for fillings of one or less
than one particle per site, as
$\chi_{\mathrm{K}}(\tau)=\chi_{\mathrm{K}}^{\mathrm{h}}(\tau)+\chi_{\mathrm{K}}^{\mathrm{d}}(\tau)+\chi_{\mathrm{K}}^{\mathrm{hd}}(\tau)$,
where
$\chi_{\mathrm{K}}^{\mathrm{h}}(\tau)$,
$\chi_{\mathrm{K}}^{\mathrm{d}}(\tau)$, and
$\chi_{\mathrm{K}}^{\mathrm{hd}}(\tau)$ are, respectively,
given as
\begin{align}
 \chi_{\mathrm{K}}^{\mathrm{h}}(\tau)
 &=NJ^2\sum_{\langle{i,j}\rangle}
   \mathcal{\bar{G}}^{\mathrm{b}}_{\mathrm{at}}(\tau)
   \mathcal{\bar{G}}^{\mathrm{b}}_{\mathrm{at}}(-\tau)
   \langle{T_{\tau}h_i(\tau)h^{\dagger}_j(\tau)h_j(0)h_i^{\dagger}(0)}\rangle,
 \label{eq:KCF1}\\
 \chi_{\mathrm{K}}^{\mathrm{d}}(\tau)
 &=J^2\sum_{\sigma_1,\sigma_2,\sigma_3,\sigma_4}\sum_{\langle{i,j}\rangle}
   \mathcal{\bar{G}}^{\mathrm{b}}_{\mathrm{at}}(\tau)
   \mathcal{\bar{G}}^{\mathrm{b}}_{\mathrm{at}}(-\tau)
 \nonumber \\
 &\quad \times
   \langle
     T_{\tau}D_{i,\sigma_1\sigma_2}^{\dagger}(\tau)D_{j,\sigma_1\sigma_3}(\tau)
             D_{j,\sigma_4\sigma_3}^{\dagger}(0)D_{i,\sigma_4\sigma_3}(0)
   \rangle,
 \label{eq:KCF2}\\
 \chi_{\mathrm{K}}^{\mathrm{hd}}(\tau)
 &=J^2\sum_{\sigma_1,\sigma_2}\sum_{\langle{i,j}\rangle}
    \biggl[
      [\mathcal{\bar{G}}^{\mathrm{b}}_{\mathrm{at}}(\tau)]^2
 \nonumber \\
 &\quad \times
      \langle{T_{\tau}h^{\dagger}_{j}(\tau)D_{i,\sigma_1\sigma_2}^{\dagger}(\tau)h_{j}(0)D_{i,\sigma_1\sigma_2}(0)}\rangle
 \nonumber \\
 &\quad
   +(\tau\rightarrow -\tau)
    \biggr].
 \label{eq:KCF3}
\end{align}
They are diagrammatically illustrated in
Fig.~\ref{fig:chi-diagrams}.
In order to simplify the form of the equations the secondary
doublon operator $D_{j,\sigma_1\sigma_2}$, which
annihilates the doublon consisting of a $\sigma_1$- and a
$\sigma_2$-component atom at site $j$, has been introduced:
\begin{equation}
 D_{j,\sigma_1\sigma_2}
 =\left\{
 \begin{aligned}
  &d_{j,\sigma_1\sigma_2} & (\sigma_1>\sigma_2) \\
  &0 & (\sigma_1=\sigma_2) \\
  &-d_{j,\sigma_2\sigma_1} & (\sigma_1<\sigma_2)
 \end{aligned}
 \right..
\end{equation}

The correlation function $\chi_{\mathrm{K}}(\tau)$ can be intuitively
interpreted as follows:
At initial time, a pair consisting of a doublon and a holon is produced
by $H_{\mathrm{K}}(0)$, and they move in the system.
Then the motion of the created doublon and holon scrambles up the spin
configuration of the initial state.
For the correlation function to be finite in the spin-incoherent case,
the spin configuration of the final state must be the same as the
initial one. The most relevant motion would thus be a retraceable path
as proposed by Brinkman and Rice~\cite{Brinkman.Rice/PRB2.1970}.
Eventually, the doublon and holon go back to the original point of the
production, and the final state created by acting
$H_{\mathrm{K}}(\tau)$ reproduces the initial state. 

\begin{figure}[tbp]
  \begin{center}
   \includegraphics[scale=1.1]{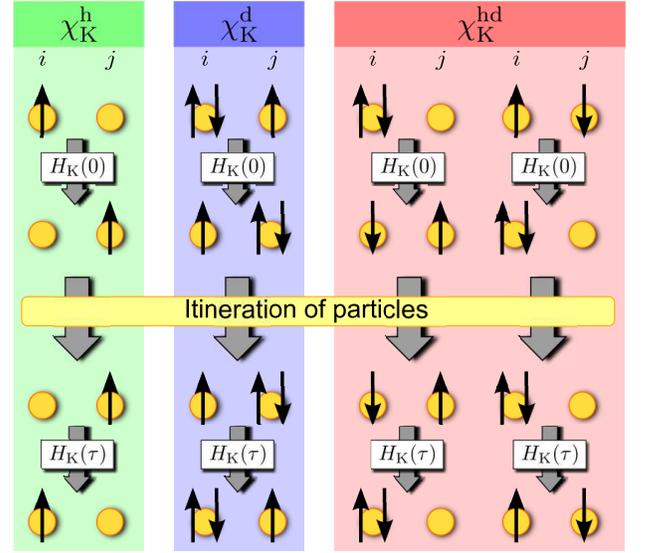}
   \caption{(Color online)
   A sketch of the contributions to the kinetic-energy correlation
   function.
   There are four possible nearest-neighboring pairs in an equilibrium
   state: (a) singly occupied and empty site, (b) doubly and singly
   occupied site, (c) singly occupied sites and (d) doubly occupied and
   empty site.
   Based on these configurations of the pair of nearest-neighboring
   sites, we can categorize the three types of contributions,
   $\chi_{\mathrm{K}}^{\mathrm{h}}(\tau)$ for pair (a) in the left
   panel, $\chi_{\mathrm{K}}^{\mathrm{d}}(\tau)$ for pair (b) in the
   middle panel, and $\chi_{\mathrm{K}}^{\mathrm{hd}}(\tau)$ for
   pairs (c) and (d) in the right panel.
   The $N=2$ case is taken here for the sake of simplicity, but it can
   easily be extended to general $N$ cases.}
   \label{fig:pairs}
  \end{center}
\end{figure}

As illustrated in Fig.~\ref{fig:pairs}, depending on the spin
configuration of the atoms in the initial state, the terms in the
correlation function, Eqs.~(\ref{eq:KCF1})--(\ref{eq:KCF3}), contribute
as follows:
$\chi_{\mathrm{K}}^{\mathrm{h}}(\tau)$ for nearest-neighboring pairs of
singly occupied and empty sites,
$\chi_{\mathrm{K}}^{\mathrm{d}}(\tau)$ for nearest-neighboring pairs of
doubly occupied and singly occupied sites,
and $\chi_{\mathrm K}^{\mathrm{hd}}(\tau)$ for nearest-neighboring pairs
of singly occupied sites and pairs of doubly occupied and empty site.
From this interpretation,
$\chi_{\mathrm{K}}^{\mathrm{h}}(\tau)$ and
$\chi_{\mathrm{K}}^{\mathrm{d}}(\tau)$ are expected to be suppressed
when the system is at a filling of one particle per site.
Thus only $\chi_{\mathrm{K}}^{\mathrm{hd}}(\tau)$ leads to important
contributions to the DPR spectrum.
In contrast, in going away from the filling of one particle per site,
the contributions of 
$\chi_{\mathrm{K}}^{\mathrm{h}}(\tau)$ and
$\chi_{\mathrm{K}}^{\mathrm{d}}(\tau)$ appear.

If one neglects vertex corrections, the two-particle correlation
functions of Eqs.~(\ref{eq:KCF1})--(\ref{eq:KCF3}) are contracted by
Wick's expansion.
As a result, the correlation functions are analytically given as
\begin{align}
 \chi_{\mathrm{K}}^{\mathrm{h}}(\tau)
 &=\mathcal{V}NzJ^2
   \mathcal{G}^{\mathrm{b}}(\tau)\mathcal{G}^{\mathrm{b}}(-\tau)
   \mathcal{G}^{\mathrm{h}}(\tau)\mathcal{G}^{\mathrm{h}}(-\tau),
 \label{eq:KCF4}
 \\
 \chi_{\mathrm{K}}^{\mathrm{h}}(\tau)
 &=\mathcal{V}\frac{N(N^3-4N+3)}{3}zJ^2
 \nonumber \\
 &\quad \times
   \mathcal{G}^{\mathrm{b}}(\tau)\mathcal{G}^{\mathrm{b}}(-\tau)
   \mathcal{G}^{\mathrm{d}}(\tau)\mathcal{G}^{\mathrm{d}}(-\tau),
 \label{eq:KCF5}
 \\
 \chi_{\mathrm{K}}^{\mathrm{hd}}(\tau)
 &=-\mathcal{V}N(N-1)zJ^2
    \biggl[
      \left[\mathcal{G}^{\mathrm{h}}(-\tau)\right]^2\mathcal{G}^{\mathrm{h}}(\tau)\mathcal{G}^{\mathrm{d}}(\tau)
 \nonumber \\
 &\quad
      +\left[\mathcal{G}^{\mathrm{h}}(\tau)\right]^2\mathcal{G}^{\mathrm{h}}(-\tau)\mathcal{G}^{\mathrm{d}}(-\tau)
    \biggr].
 \label{eq:KCF6}
\end{align}
By moving from the imaginary-time domain to the real-time one by
analytic continuation, and by taking the imaginary part of the
correlation functions, the analytic form of the DPR per site at a
filling of one or less than one particle per site 
is finally obtained as
\begin{align}
 P_{\mathrm{D}}(\omega)
 &=\frac{(\delta{F})^2\omega}{\hbar U}
   \int\!\!\frac{d\nu}{2\pi}\
   \biggl[
     \biggl(f(\hbar\nu-\hbar\omega)-f(\hbar\nu)\biggr)
 \nonumber \\
 &\quad \times
     \biggl\{
       \frac{W_{\mathrm{h}}^{2}}{8}
       \mathcal{A}^{\mathrm{h}}(\nu)
       \mathcal{A}^{\mathrm{h}}(\nu-\omega)
       +\frac{N(N^2+N-3)}{6}\frac{W_{\mathrm{d}}^2}{8}
 \nonumber \\
 &\quad \times
        \mathcal{A}^{\mathrm{d}}(\nu)
        \mathcal{A}^{\mathrm{d}}(\nu-\omega)
     \biggr\}
     \nonumber \\
     &\quad
     +\frac{N(N-1)zJ^{2}\left(1+2b(\epsilon^{\mathrm{b}})\right)}{2}
        \biggl(f(\hbar\nu)+b(2\epsilon^{\mathrm{b}})\biggr)
 \nonumber \\
 &\quad \times
        \biggl(f(2\epsilon^{\mathrm{b}}-\hbar\nu)-f(2\epsilon^{\mathrm{b}}-\hbar\nu+\hbar\omega)\biggr)
 \nonumber \\
 &\quad \times
        \mathcal{A}^{\mathrm{h}}(\nu)
        \mathcal{A}^{\mathrm{d}}(2\epsilon^{\mathrm{b}}/\hbar-\nu+\omega)
 \nonumber \\
 &\quad
     +\frac{N(N-1)zJ^{2}\left(1+2b(\epsilon^{\mathrm{b}})\right)}{2}
        \biggl(f(\hbar\nu)+b(2\epsilon^{\mathrm{b}})\biggr)
 \nonumber \\
 &\quad \times
        \biggl(f(2\epsilon^{\mathrm{b}}-\hbar\nu-\hbar\omega)-f(2\epsilon^{\mathrm{b}}-\hbar\nu)\biggr)
 \nonumber \\
 &\quad \times
        \mathcal{A}^{\mathrm{h}}(\nu)
        \mathcal{A}^{\mathrm{d}}(2\epsilon^{\mathrm{b}}/\hbar-\nu-\omega)
   \biggr].
 \label{eq:DPR_spectrum}
 \end{align}

\begin{figure*}[tbp]
 \begin{center}
  \includegraphics[scale=.8]{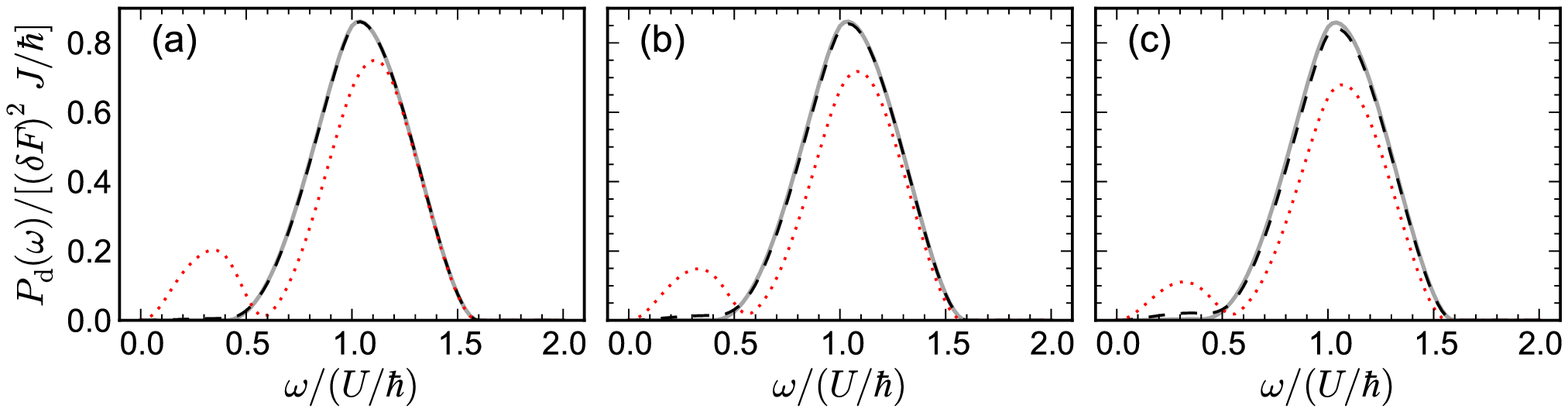}
  \caption{(Color online)
  The DPR spectra per site,
  $P_{\mathrm{D}}(\omega)/[(\delta{F})^2J/\hbar]$, as a
  function of modulation frequency for $N=2$ in a cubic lattice ($z=6$):
  (a) $k_{B}T/U=0.025$ ($k_{B}T < J$),
  (b) $k_{B}T/U=0.05$ ($k_{B}T = J$), and
  (c) $k_{B}T/U=0.075$ ($k_{B}T > J$).
  The hopping parameter is taken to be $J/U=0.05$.
  The solid, dashed, and dotted lines denote $\mu/U=0.5$, $0.3$, and
  $0.1$, respectively.
  }
  \label{fig:DPR-spectra-N2}
 \end{center}
 \begin{center}
  \includegraphics[scale=.8]{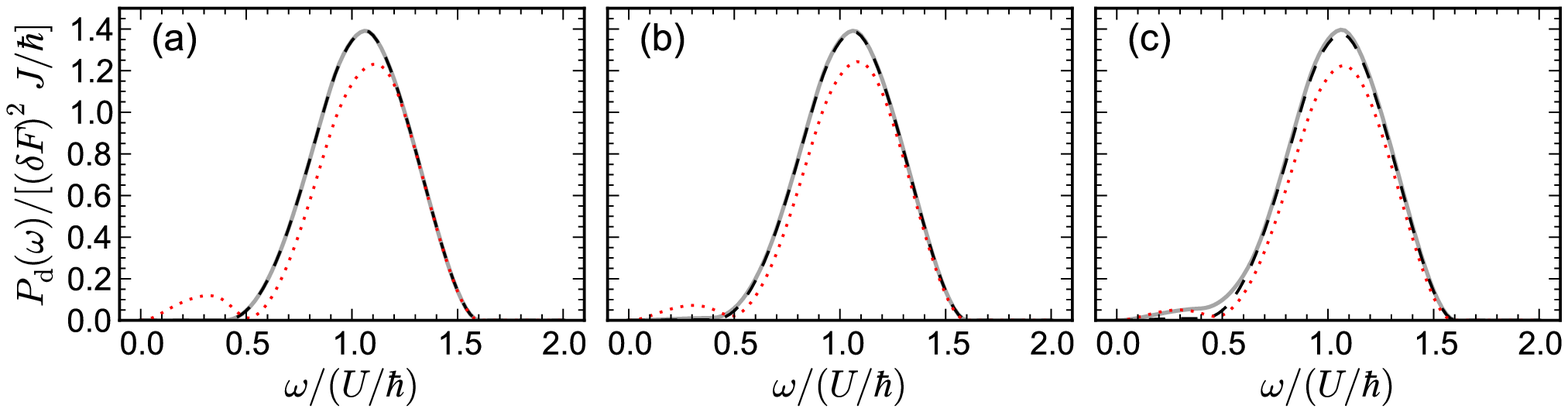}
  \caption{(Color online)
  The DPR spectra per site,
  $P_{\mathrm{D}}(\omega)/[(\delta{F})^2J/\hbar]$, as a
  function of modulation frequency for $N=6$ in a cubic lattice ($z=6$):
  (a) $k_{B}T/U=0.025$ ($k_{B}T < J$),
  (b) $k_{B}T/U=0.05$ ($k_{B}T = J$), and
  (c) $k_{B}T/U=0.075$ ($k_{B}T > J$).
  The hopping parameter is taken to be $J/U=0.05$.
  The solid, dashed, and dotted lines denote $\mu/U=0.5$, $0.3$, and
  $0.1$, respectively.
  }
  \label{fig:DPR-spectra-N6}
 \end{center}
 \begin{center}
  \includegraphics[scale=.8]{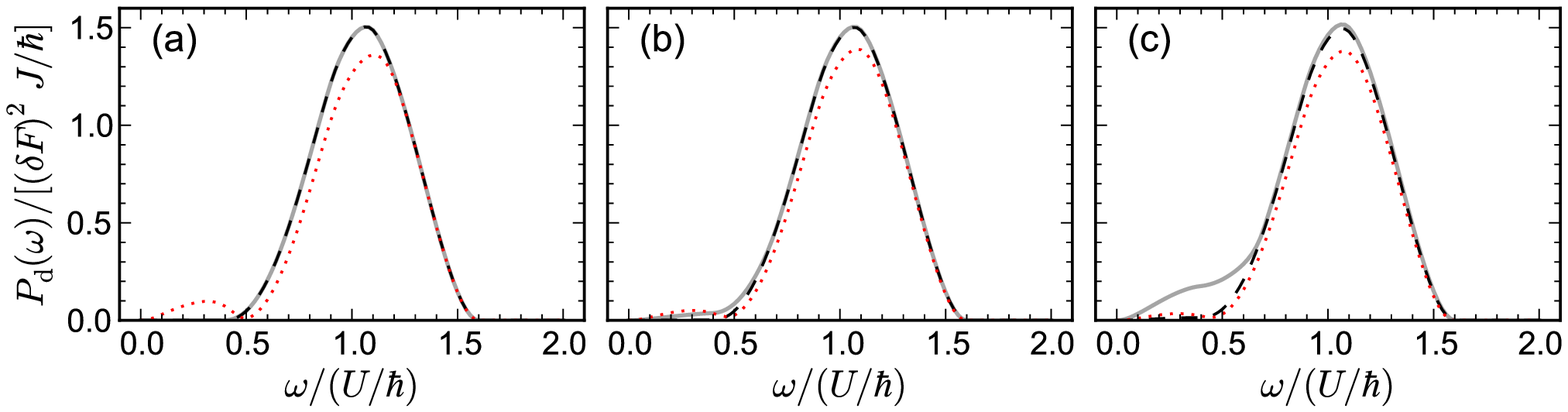}
  \caption{(Color online)
  The DPR spectra per site,
  $P_{\mathrm{D}}(\omega)/[(\delta{F})^2J/\hbar]$, as a
  function of modulation frequency for $N=10$ in a cubic lattice
  ($z=6$):
  (a) $k_{B}T/U=0.025$ ($k_{B}T < J$),
  (b) $k_{B}T/U=0.05$ ($k_{B}T = J$), and
  (c) $k_{B}T/U=0.075$ ($k_{B}T > J$).
  The hopping parameter is taken to be $J/U=0.05$.
  The solid, dashed, and dotted lines denote $\mu/U=0.5$, $0.3$, and
  $0.1$, respectively.
  }
  \label{fig:DPR-spectra-N10}
 \end{center}
\end{figure*}

The DPR spectra for different $N$s and chemical potentials
are shown in Figs.~\ref{fig:DPR-spectra-N2}--\ref{fig:DPR-spectra-N10},
where the small hopping is fixed to be $J/U=0.05$.
To illustrate the temperature dependence, the spectra for
$k_{B}T/U=0.025$ ($<J/U$), $0.05$ ($=J/U$), and $0.075$ ($>J/U$) are
given.
As expected, the dominant peak is found to appear around
$\omega=U/\hbar$ for every $\mu/U$ and $N$, and the peak becomes sharper
as $\mu/U$ gets closer to $1/2$, and temperature is lowered.
Away from filling unity, another small peak in the lower frequency
regime appears.
It occurs because $\chi_{\mathrm{K}}^{\mathrm{h}}$ becomes relevant due
to the hole doping.
The spectral weight of this small peak away from filling unity, 
e.g., at $\mu/U=0.1$, tends to be suppressed for any
$N$ as temperature increases.
As shown in Fig.~\ref{fig:DPR-spectra-N6}
and~\ref{fig:DPR-spectra-N10}, the weight of the peak around
$\omega=U/\hbar$ for $N>2$ increases with $N$.
This is due to the enhancement caused by the larger spectral weight
shown in
Figs.~\ref{fig:spectralfunctionN=2}--\ref{fig:spectralfunctionN=10}
as $N$ increases.
For such $N$s, the spectral weight in the low-frequency regime away from
$\mu/U=0.5$, which comes from $\chi_{\mathrm{K}}^{\mathrm{h}}$, is
suppressed, in contrast to what happens for the $N=2$ case.
However, interestingly, the spectral weight in the low-frequency
regime for $N=6$ and $10$ also increases with temperature,
while this strong tendency is not seen in the case of $N=2$.
This is due to the finite contribution of
$\chi_{\mathrm{K}}^{\mathrm{d}}$ because the doublon band enhanced by
the larger $N$ reaches $\omega=0$ in such a parameter regime. This is
shown in Figs.~\ref{fig:spectralfunctionN=6}
and~\ref{fig:spectralfunctionN=10}. 

\begin{figure}[htbp]
 \begin{center}
  \includegraphics[scale=1.]{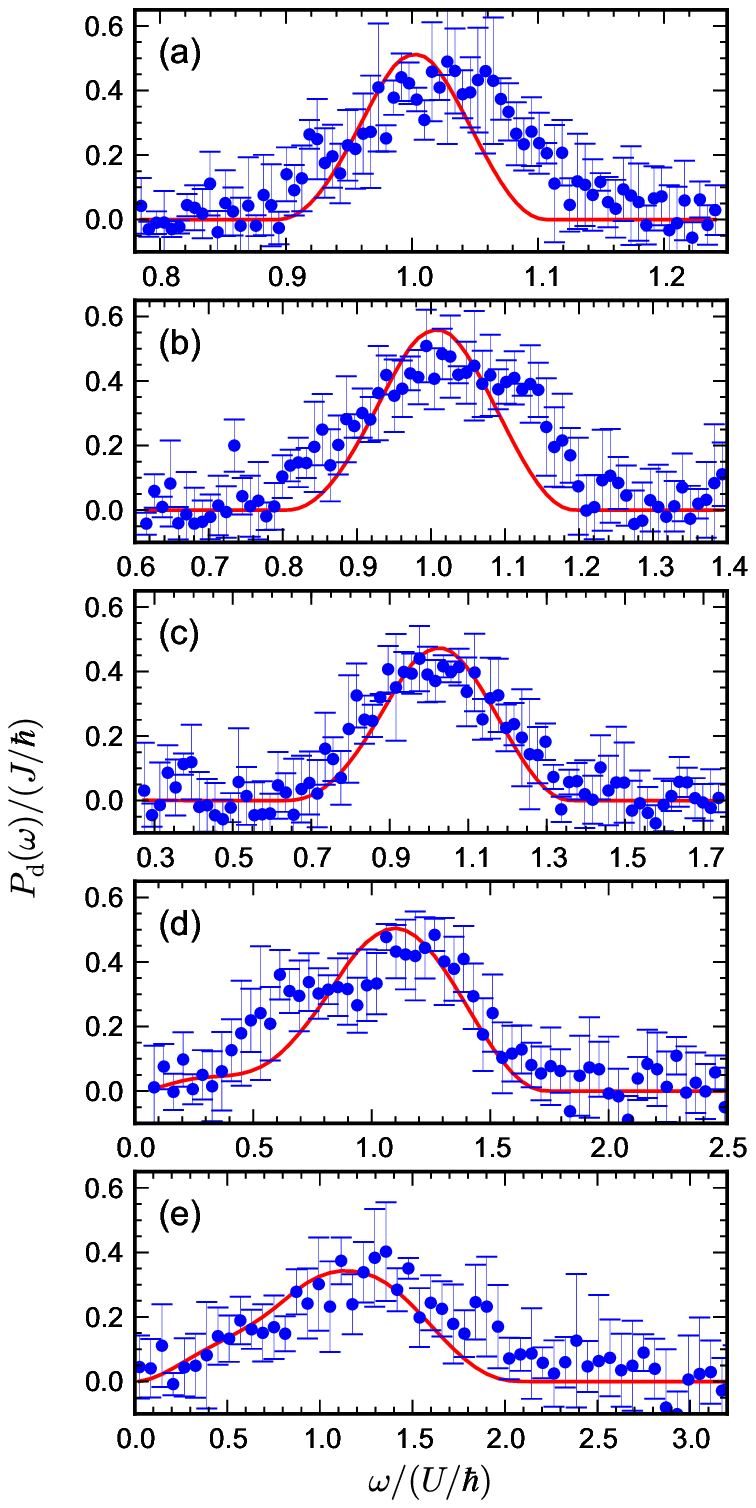}
  \caption{(Color online)
  The dimensionless DPR spectra $P_{\mathrm{d}}(\omega)$ scaled by
  $J/\hbar$ as a function of modulation frequency $\hbar\omega/U$ for an
  $N=6$ trapped system with different parameters: (a) $J/U=0.0089$, (b)
  $J/U=0.016$, (c) $J/U=0.030$, (d) $J/U=0.061$, and (e) $J/U=0.091$.
  In all cases, $1.87 \times 10^4$ trapped atoms and a
  modulation amplitude $\delta{F}=0.08$ are taken.
  The solid line and points with an error bar, respectively, denote the
  theoretical result and the $^{173}$Yb
  experiments~\cite{Taie.etal/NatPhys2012}.
  The temperatures are determined by the least-square fit to the
  experiment, and the values are (a) $k_{B}T/U=0.0719$,
  (b) $k_{B}T/U=0.0577$, (c) $k_{B}T/U=0.0909$, (d) $k_{B}T/U=0.0963$,
  and (e) $k_{B}T/U=0.179$.
  }
  \label{fig:exp-fit}
 \end{center}
\end{figure}

In addition to the plots of
Figs.~\ref{fig:DPR-spectra-N2}--\ref{fig:DPR-spectra-N10}, we also
directly fit our results to the
experiment~\cite{Taie.etal/NatPhys2012}. 
In such experiments, done with $^{173}$Yb atoms, the system is
expected to be dominated by the MI, and thus our calculation scheme at a
filling of one or less than one particle per site would be applicable,
using an LDA calculation to take the trap into account.
The results, using our theoretical analysis, when taking the
parameters corresponding to the experiment are shown 
in Fig.~\ref{fig:exp-fit}.
The experimental parameters, namely, the hopping energy $J/U$, the trap
frequency, and the modulation amplitude $\delta{F}$, are taken as
follows:
(a) $J/U=0.0089$, $2\pi\times(172,139,64)$ [Hz], and $\delta{F}=0.085$,
(b) $J/U=0.016$, $2\pi\times(175,141,69)$ [Hz], and $\delta{F}=0.09$,
(c) $J/U=0.030$, $2\pi\times(181,146,78)$ [Hz], and $\delta{F}=0.10$,
(d) $J/U=0.061$, $2\pi\times(187,151,86)$ [Hz], and $\delta{F}=0.115$,
and (e) $J/U=0.091$, $2\pi\times(193,155,86)$ [Hz], and
$\delta{F}=0.125$, respectively.
The number of atoms in the trap is commonly assumed to be
$1.87\times10^4$. 
The following temperatures are determined by the least-square fits to
the experimental data:
(a) $k_{B}T/U=0.0719$,
(b) $k_{B}T/U=0.0577$,
(c) $k_{B}T/U=0.0909$,
(d) $k_{B}T/U=0.0963$,
and (e) $k_{B}T/U=0.179$.
Figure~\ref{fig:exp-fit} shows good agreement with the experimental
result, which supports the validity of our theory.

\section{Summary}~\label{sec:summary}
We have computed doublon and holon excitations of strongly interacting
$N$-component fermions in optical lattices in the spin-incoherent regime.
This corresponds to a temperature region between the superexchange
coupling and the interaction.
As an effective Hamiltonian to extract the physics at an energy scale of
order $\sim U$, the symmetric
SU($N$) Hubbard model has been studied, which means that the Hubbard
interaction is independent of the internal degree of freedom of the
fermions. 
The theory presented in Ref.~\cite{Tokuno.Demler.Giamarchi/PRA85.2012},
which reproduces well the experiment with $^{40}$K
atoms~\cite{Greif.etal/PRL106.2011}, has 
been extended to an $N$-component fermion case, and the analytic form of
the single-particle spectral functions for fillings of one or less than
one particle per site has been obtained.
Our approach is based on the slave-particle representation
in which the original fermion operators are represented by a fermionic
holon, $N$ species of bosonic spinons, and $N(N-1)/2$ species of
fermionic doublons.
We have employed a combination of mean-field theory,
a diagrammatic approach, and the NCA to take into account the
finite particle hopping $J$, and we have captured the physics of the
hole-doped systems for large interaction $J\ll U$.

As an application to the calculation of the experimental observables,
the DPR induced by dynamical periodic modulation of optical lattices as
a function of modulation frequency has been also computed, both for the
homogeneous system and for the trapped system, in an LDA.
As shown in the Appendix, the DPR spectrum as a second-order response to
the optical lattice modulation is directly related to the retarded
kinetic-energy correlation function. 
We have discussed the DPR spectrum without vertex corrections, and we
have presented the analytic form constructed by the obtained spectral
functions of the doublon and the holon.

From the obtained analytic form in the case of homogeneous systems, we
have obtained the DPR spectra as a function
of temperature, chemical potential, and component number $N$, and we
have compared the different behaviors for the different $N$s.
While the large peak structure around the interaction $U$ exists
regardless of the value of $N$, some differences have been observed
in the regime of low modulation frequency.
In the comparison, we have focused on two different effects leading to
an enhancement of the spectral weight in the low-frequency regime.
The first one is a doping effect: in going away from half-filling, the
low-frequency spectrum appears as a consequence of the system
becoming metallic.
This effect has been found to be suppressed as $N$ increase.
The second effect is the temperature: the spectral weight in the
low-frequency side tends to increase with temperature.
However, unlike the first effect, we find that the spectral weight
is enhanced as $N$ goes up.
Therefore the properties of the spectra for different $N$s will be most
markedly different for the low-energy part of the spectrum.

The theory presented in this paper has several advantages: First, the
finite-temperature dynamics can be dealt with analytically.
For such dynamical correlations, numerical
approaches cannot be straightforwardly applied because of the difficulty
of numerical implementation of the analytic continuation;
second, our theoretical technique allows for the control of the chemical
potential, in principle. Note however that our approximations are
expected to work well at a filling close to the MI state.
This means that inhomogeneous systems in the presence of a trap
potential can be also discussed by using an LDA.
Indeed, in Ref.~\cite{Tokuno.Demler.Giamarchi/PRA85.2012}, this
approach has been applied to the SU($2$) symmetric Hubbard model with a
harmonic trap potential, and quantitatively precise agreement with the
experiment~\cite{Greif.etal/PRL106.2011} has been obtained.

Using the extension of this approach to trapped systems we have compared
our results for the DPR spectra shown in Fig.~\ref{fig:exp-fit}, for
which the presence of the trap potential is taken into account by LDA,
with $^{173}$Yb experiments.
The temperature has been determined by the best fit to the experimental
data~\cite{Taie.etal/NatPhys2012}, and the obtained results for the DPR
peak are in good agreement with the experiment.

In recent years, the symmetric SU($N$) systems have been being realized
in experiments with alkaline-earth-metal(-like)
$^{87}$Sr~\cite{Ye.et.al/Science320.2008,DeSalvo.et.al/PRL105.2010,Tey.etal/PRA82.2010}
in addition to $^{173}$Yb atoms.
Current fermionic atom systems in such experiments are still at high
temperature.
Therefore our theory is expected to work very effectively to compare
up-coming lattice modulation experiments in such a temperature regime.

Finally, we would like to mention some prospects of our study.
The first is to apply this technique to the calculation of thermodynamic
functions such as entropy.
It is hard to measure temperature directly in experiments, and the
measurement of entropy is used instead.
Thus by computing the entropy within our theoretical framework, we can
make a more straightforward comparison with the experiment.
The second is to extend the theory to general $N$-component
mixtures away from the SU($N$) symmetry limit.
Although the SU($N$) symmetry has been assumed throughout this paper,
the slave-particle representation and the NCA calculation would be still
applicable away from the SU($N$) symmetric point.
However, the self-consistent equations for the
self-energies [Eqs.~(\ref{eq:Holon_self-energy})
and~(\ref{eq:Doublon_self-energy})]
remain complicated, and the issue would be how to solve the
self-consistent equations.
Another prospect is to develop this technique to capture the
low-temperature physics.
The difficulty of the application of the present technique to
spin-coherent systems is that we have here assumed fully incoherent
spins.
Namely, the spinon propagators are replaced by the atomic propagators,
which means that even nearest-neighbor spin correlations are ignored.
Thus the key to improve the technique for lower temperature would be to
modify the spin-incoherence
assumption~(\ref{eq:Spinon_propagator_replacement}) .
Such an improved technique would allow for the crosscheck of the theoretical
predictions~\cite{Bonnes.etal/arXiv2012,Messio.Mila/arXiv2012}.

\begin{acknowledgements}
 We thank A. Lobos for a fruitful discussion on the slave-particle
 method and S. Taie and Y. Takahashi for valuable intensive discussions
 on experiments of $^{173}$Yb atoms.
 This work was supported by the Swiss National Foundation under MaNEP
 and division II.
\end{acknowledgements}

\appendix
\section{Formulation of DPR spectra}\label{DPRformalism}
The derivation of the DPR formula is briefly reviewed.
For simplicity, we only consider the homogeneous case, but the more
general case including an inhomogeneous potential such as a trap can be
discussed.
Such a general argument can be found in
Ref.~\cite{Tokuno.Giamarchi/PRA85.2012}.
We start with a generic Hamiltonian of interacting atoms in optical
lattice potentials defined in $D$-dimensional continuum space, which is
written as follows:
\begin{equation}
 H=H_{0} + \int\!\! d\bm{r} V_{\rm op}(\bm{r})\rho(\bm{r}),
\end{equation}
where $V_{\rm op}(\bm{r})=\sum_{\mu=1}^{D}V_0\cos^2(k x_{\mu})$
is the optical lattice potential, and $H_{0}$ is an unperturbed
Hamiltonian of interacting atoms in free space.

For a deep optical lattice potential ($V_0\gg\mu$), the Hamiltonian
$H$ is well described by the Hubbard model.
Then the parameters, the hopping $J$ and on-site interaction $U$,
are given as a function of lattice depth $V_0$.
For example, if the Wannier function is assumed to be approximated as a
Gaussian wave function, the hopping $J$ and on-site interaction $U$ are
estimated~\cite{Bloch.Dalibard.Zwerger/RevModPhys80.2008} as
\begin{align}
 J
 &\approx
 \frac{4}{\sqrt{\pi}}E_{\mathrm{R}}\left(\frac{V_0}{E_{\mathrm{R}}}\right)^{3/4}
 \exp\left[-2\sqrt{\frac{V_0}{E_{\mathrm{R}}}}\right],
 \label{eq:hopping}\\
 U
 &\approx
 \frac{8}{\pi}ka_{\mathrm{s}}E_{\mathrm{R}}
 \left(\frac{V_0}{E_{\mathrm{R}}}\right)^{3/4},
 \label{eq:interaction}
\end{align}
where $E_{\mathrm{R}}$ and $a_{\mathrm{s}}$ are, respectively, the
recoil energy and $s$-wave scattering length of atoms in free space.

We consider an amplitude modulation perturbation of an optical
lattice. 
For deep lattices the modulation effect of the lattice potential can be
described by replacing the amplitude of the static lattice potential
$V_{\mathrm{op}}$ as $V_0\rightarrow V_0[1+\delta{V}\cos(\omega t)]$.
Then the parameters of the lattice model also follow the replacement,
and the modulation parts are derived up to first order in
$\delta{V}$:
\begin{align}
 U&\rightarrow U[1+\delta{U}\cos(\omega t)],
 \\
 J&\rightarrow J[1+\delta{J}\cos(\omega t)],
\end{align}
where
$\delta{U}=(\partial{\ln{U}}/\partial{V_0})V_0$ and
$\delta{J}=(\partial{\ln{J}}/\partial{V_0})V_0$.
In the case of a Gaussian Wannier function~(\ref{eq:hopping})
and~(\ref{eq:interaction}), $\delta{U}\approx 3/4$ and
$\delta{J}\approx 3/4-\sqrt{V_0/E_{\mathrm{R}}}$.
Thus the time-dependent perturbation by lattice modulation is written as
$(\delta{U}H_{\mathrm{U}}+\delta{J}H_{\mathrm{K}})\cos(\omega t)$, where
$H_{\mathrm{U}}=U\sum_{j,\sigma_1>\sigma_2}n_{j,\sigma_1}n_{j,\sigma_2}$.
In addition, by making use of the form of the Hubbard Hamiltonian, the
perturbation can be rewritten as
$\delta{U}H\cos(\omega t)+\delta{F}H_{\mathrm{K}}\cos(\omega t)$, where
$\delta{F}=\delta{J}-\delta{U}$.
Thus the considered Hamiltonian with the lattice modulation is written as
\begin{equation}
 H(t)=H+\delta{U}H\cos(\omega t)+\delta{F}H_{\mathrm{K}}\cos(\omega t).
 \label{eq:Hamiltonian2}
\end{equation}

Extending the doublon number projector for $N=2$, we can define the
total number operator of a doublon by the Hubbard interaction as
\begin{equation}
 N_{\rm D}=\frac{1}{U}H_{\mathrm{U}},
  \label{eq:DNP}
\end{equation}
where we have defined the total doublon number as a total sum of all
species of doublons.
This projection operator~(\ref{eq:DNP}) truncates only the empty and
singly occupied state, and thus it could not count the doublon number
perfectly for $N>2$ since the projected states include
multiparticle occupancies of more than three such as three- and four-fold
occupancy and so on.
However, because such multiparticle occupations are away from the
main energy scale in the case considered here , Eq.~(\ref{eq:DNP})
would be also identical to the total doublon number in the case of
multicomponent fermions.
The DPR per site is defined as a time average of time derivative of
$N_{\mathrm{D}}$ over a single period of modulation:
\begin{equation}
 P_{\mathrm{D}}(\omega)
 \equiv \frac{1}{2\pi/\omega}
        \int_{T}^{T+2\pi/\omega}\!\! dt\
         \frac{d}{dt}\frac{\langle{N}_{\mathrm{D}}\rangle}{\mathcal{V}},
\end{equation}
where $\langle{\cdots}\rangle$ means the thermodynamic average by the
Hamiltonian (\ref{eq:Hamiltonian2}), and $\mathcal{V}$ is the total
number of lattice sites.

We implement second-order perturbation theory in term of
$d\langle{N_{\mathrm{D}}}\rangle/dt$.
Then we use the following mathematical trick: 
Using Eq.~(\ref{eq:Hamiltonian2}), we rewrite the total doublon number
operator as
\begin{equation}
 N_{\mathrm{D}}
 =\frac{1}{U}
    \biggl[H(t)-H_{\mathrm{K}}-(H+\delta{F}H_{\mathrm{K}})\cos(\omega t)
  \biggr].
\end{equation}
From a straightforward calculation up to second order, the terms
apart from $H(t)$ are found to contribute as oscillatory terms, and
they cancel due to the single-period time average.
Thus $P_{\mathrm{D}}(\omega)$ can be rewritten as
\begin{equation}
 P_{\mathrm{D}}(\omega)
 \equiv \frac{1}{2\pi/\omega}
        \int_{T}^{T+2\pi/\omega}\!\! dt\
         \frac{\langle{\dot{H(t)}}\rangle}{\mathcal{V}},
\end{equation}
where we have used the identity
$d\langle{H(t)}\rangle/dt=\langle{\dot{H}(t)}\rangle$.
This is equivalent to the definition of the energy absorption
rate~\cite{Tokuno.Gimarchi/PRL106.2011}.
This equivalence was numerically established for spin-$1/2$ one-dimensional fermions in Ref.~\cite{Kollath.etal/PRA74.2006}.
The second-order response of the energy absorption rate can be
calculated by linear response.
Therefore one can finally obtain the formula as
\begin{equation}
 P_{\rm D}(\omega)
 =-\frac{(\delta{F})^2}{2\hbar\mathcal{V}U}\omega\,
   \mathrm{Im}\,{\tilde{\chi}^{\mathrm{R}}_{\mathrm{K}}(\omega)},
 \label{eq:generalDPR}
\end{equation}
where $\tilde{\chi}^{\rm R}_{\mathrm{K}}(\omega)$ is a Fourier
component of the kinetic-energy-retarded correlation function
$\chi^{\mathrm{R}}_{\mathrm{K}}(t)=-i\theta(t)\langle[H_{\mathrm{K}}(t),H_{\mathrm{K}}(0)]\rangle_0$
where $\langle{\cdots}\rangle_0$ denotes the statistical average by the
unperturbed Hamiltonian $H$.

\bibliographystyle{apsrev4-1}
%

\end{document}